\documentclass[prl,twocolumn,amsmath,superscriptaddress,amssymb,showkeys,10pt]{revtex4-2}

\usepackage{color}
\usepackage{graphicx}
\usepackage{epstopdf}
\usepackage{bm}
\usepackage{gensymb}
\usepackage{amsmath}       
\usepackage{ulem}       

\usepackage[colorlinks=true,urlcolor=blue,linkcolor=blue,citecolor=blue]{hyperref}
\usepackage{hyperref}       

\renewcommand{\thesection}{\Roman{section}}       
\renewcommand{\thesubsection}{\Alph{subsection}}  

\begin{document}

\DeclareGraphicsExtensions{.eps, .png, .pdf}

\def\SrFour{Sr$_4$Ru$_3$O$_{10}$}
\def\SrThree{Sr$_3$Ru$_2$O$_7$}
\def\SrTwo{Sr$_2$RuO$_4$}
\def\SrOne{SrRuO$_3$}
\def\CaOne{CaRuO$_3$}
\def\SrSeries{Sr$_{n+1}$Ru$_n$O$_{3n+1}$}
\def\FGT{Fe$_3$GeTe$_2$}

\def\etal{$et\ al.$}

\def\EF{$E_\mathrm{F}$}
\def\EFzero{$E_\mathrm{F}^0$}
\def\kF{$k_\mathrm{F}$}
\def\vF{$v_\mathrm{F}$}
\def\kpar{$k_\mathrm{||}$}
\def\kB{$k_\mathrm{B}$}
\def\Tcoh{$T^*$}
\def\invA{\AA$^{-1}$}
\def\kpar{$k_{\vert\vert}$}
\def\kz{$k_z$}
\def\pz{$p_z$}
\def\deg{$^{\circ}$}
\def\kxky{$k_x$-$k_y$ }
\def\kxkz{$k_x$-$k_z$ }
\def\hv{$h\nu$}
\def\Akw{$A(k,\omega)$}
\def\dxy{$d_{xy}$}
\def\dxzyz{$d_{xz/yz}$}

\def\G{$\Gamma$}
\def\p{$\pi$}
\def\LW{$\gamma_L$}
\def\GW{$\gamma_G$}
\def\VW{$\gamma_V$}


\def\cred{\color{red}}
\def\cblue{\color{blue}}
\definecolor{dkgreen}{rgb}{0.15,0.45,0.10}
\def\cgreen{\color{dkgreen}}
\def\cmag{\color{magenta}}

\title{
Layer-dependent spin-resolved electronic structure\\ of
ferromagnetic triple-layered ruthenate Sr$_4$Ru$_3$O$_{10}$}

\author{Prosper Ngabonziza}
\email[]{pngabonziza@lsu.edu}
\affiliation{Department of Physics and Astronomy, Louisiana State University, Baton Rouge, Louisiana 70803, USA}
\affiliation{Department of Physics, University of Johannesburg,
P.O. Box 524 Auckland Park 2006, Johannesburg, South Africa}

\author{Jonathan D. Denlinger}
\email[]{jddenlinger@lbl.gov}

\author{Alexei V. Fedorov}
\affiliation{Advanced Light Source, Lawrence Berkeley National Laboratory, Berkeley, California 94720, USA}

\author{Gang Cao}
\affiliation{Department of Physics, University of Colorado at Boulder, Boulder, Colorado 80309, USA}

\author{J. W. Allen}
\affiliation{Randall Laboratory of Physics, University of Michigan, Ann Arbor, Michigan 48109, USA}

\author{G. Gebreyesus}  
\affiliation{Department of Physics, School of Physical and Mathematical Sciences,
College of Basic and Applied Sciences, University of Ghana, Ghana}

\author{Richard M. Martin}
\affiliation{Department of Physics, University of Illinois at Urbana-Champaign,Urbana, Illinois 61801, USA}
\affiliation{Department of Applied Physics, Stanford University, Stanford, California 94305, USA}

\date{\today}


\begin{abstract}

High-resolution angle- and spin-resolved photoemission spectroscopy (ARPES) of the triple-layered ruthenate Sr$_4$Ru$_3$O$_{10}$  reveals features of the electronic structure that extend our understanding of the layered strontium ruthenates.
The spectra near the Fermi energy are very different from the non-magnetic analogues Sr$_2$RuO$_4$ and Sr$_3$Ru$_2$O$_7$ with distinct Fermi surfaces for wide electron-like minority spin bands around the zone center and narrow hole-like majority spin Fermi surface contours around the zone corners.
The most dramatic results are two narrow spectral peaks $\sim$30 meV below the Fermi-level, a spin-minority hole-like band at the Brillouin zone center, and a spin-majority saddle-band van Hove singularity at the zone edge, which exhibits almost 100\% spin-polarization at low temperature,
and a strong temperature dependent coherence-incoherence crossover attributed to Hund metal correlations.
Quantitative comparison of the ARPES to spin-polarized density functional theory (DFT) calculations identify the specific antibonding and nonbonding orbital origins of the narrow bands, with a prediction of different spatial localization in the central and outer layers. This is shown to be consistent with experimental ARPES multi-zone matrix element intensity variations, and implicates outer-layer-specific control of the in-plane metamagnetism.
The renormalization of the bands relative to the mean-field DFT,
the demonstration of spin-polarized oxygen bands, and of spin-minority and spin-majority band-crossing hybridization, provide a more complete picture of the magnetism which displays aspects of both delocalized and local moment behavior. 
\end{abstract}

\maketitle

\tableofcontents


\section{Introduction} \label{intro}   

Strontium ruthenates of the Ruddlesden-Popper phases Sr$_{n+1}$Ru$_n$O$_{3n+1}$ ($n$ = 1, 2, 3, \dots $\infty$) play a pivotal role in studies of strongly correlated electron systems.
Depending on the number $n$ of the RuO$_6$ octahedra layers in the unit cell,
they exhibit phenomena including unconventional  superconductivity in \SrTwo\ ($n$=1)
\cite{Maeno1994},
quantum critical metamagnetism and nematic fluid with heavy $d$-electron masses in \SrThree\ ($n$=2) \cite{Putatunda2020,Tamai2008}, anisotropic ferromagnetism and in-plane metamagnetism in \SrFour\ ($n$=3) \cite{Crawford2002,Cao2003,Gupta2006,Zheng2018},
control of correlated phases in heteroepitaxial single-atomic-layer ruthenates \cite{Kim2023}, and many others.
The limit of $n$ =$\infty$ is the three-dimensional ferromagnetic crystal \SrOne.
It has often been considered to be an itinerant ferromagnet \cite{Koster2012},
but there is evidence of localized correlated behavior \cite{Shai2013,Dang2015,Hahn2021}
This rich array of distinct collective phenomena can be traced to the competition between local and itinerant behaviors,
effects of dimensionality, structural distortions, orbital polarization, crystal field splitting and spin-orbit coupling \cite{Malvestuto2011}.


 The electronic structure of the single-layer  \SrTwo\  and double-layer \SrThree\ have been studied in great detail \cite{Damascelli2000,Ingle2005,Tamai2019,Tamai2008,Allan2013}.
 The Ru ions are formally Ru$^{+4}$ with four electrons in the $4d$ shell that hybridize with the oxygen $p$ states to form quasi-two-dimensional $t_{2g}$ bands, with $d_{xz}$, $d_{yz}$ and $d_{xy}$ all partially occupied.
 The Fermi surfaces are well-described by density functional calculations, but there are strong effects of correlation which are revealed by large renormalization of the masses of bands at the Fermi energy \cite{Tamai2019,Tamai2008}.
 In \SrThree\ angle-resolved photoemission spectroscopy (ARPES) experiments \cite{Tamai2008} have revealed shallow renormalized bands with a complex density of states with van Hove singularities (vHS) near the Fermi level, a situation which is favorable for magnetic instabilities.

The strontium ruthenates are often considered to exemplify the class of materials termed Hund metals in which there are multiple bands and correlation is due primarily to intra-atomic exchange interactions characterized by the Hund's rule coupling which favors high spin configurations \cite{Georges2013,Medici2011,Deng2019,Georges2024}.
 The effects depend upon the occupancies of the atomic-like orbitals and lead to correlation on multiple energy scales.  The striking effects of correlation were established already in the 1960's for single-impurities with spin $S$ coupled to a sea of conduction electrons, where the characteristic Kondo temperature decreases exponentially with $S$ \cite{Schrieffer1967,Blandin1968,Jayaprakash1981}.
More recently it has been realized that the effects extend to crystals with bandwidth significantly larger than
the interaction 
energies as shown in a number of studies using dynamical mean field theory (DMFT),
for example, a recent calculation for  \SrTwo\ which finds a two-stage process of screening orbital and spin fluctuations, so that Fermi-liquid behavior sets in only with spin coherence below T$_{FL}$ $\sim$25 K \cite{Kugler2020}.

Triple-layer \SrFour\ adds a new dimension to correlated electron phenomena of the layered ruthenates 
with its mix of ferromagnetism and delicate metamagnetism.
Although ferromagnetic \SrOne\ has been studied extensively, \SrFour\ is a layered material with many features in common with the other layered ruthenates and, as shown in the present work, has much more dramatic effects on the electronic properties than is found in \SrOne\ \cite{Hahn2021}. 
\SrFour\ exhibits ferromagnetic order below $T_c$$\approx$105 K with a low temperature  ($T$) average moment of 1.1 $\mu_B$/Ru parallel to the $c$ axis that is about half the maximum moment of 2 for Ru$^{+4}$ ions populating three energy levels.

For weak in-plane fields, the magnetic moment rises upon cooling below $T_c$ to a maximum at $T$*$\sim$60 K and then anomalously reduces again to zero at low temperature; the suppressed low temperature moment is then enhanced with higher magnetic fields with a sharp metamagnetic transition to a large moment at $H_c$=2.5 T \cite{Cao2003,Gupta2006,Mao2006,Xu2007}.
The observation of two field-dependent steps for this in-plane metamagnetic transition in \SrFour\ \cite{Carleschi2014} has prompted speculation about the role of multiple van-Hove singularities near \EF\ and/or the inequivalence of the Ru sites in the central versus two outer layers of the tri-layer crystal structure.
Subsequent analysis of neutron diffraction data indeed identifies layer-dependent moments, and models of layer-dependent spin orientations are proposed \cite{Granata2016,Forte2019,Capogna2020}.
Also the metamagnetic behavior is very pressure sensitive \cite{Gupta2006,Zheng2018} and 4\% Ir substitution of Ru has recently been shown to stabilize the $c$-axis spin orientation and enhance the outer layer RuO$_6$ octahedra rotation angle \cite{Ye2023}.  At present there is no settled microscopic understanding of this delicate metamagnetic behavior.

Here we present spin-resolved ARPES data for \SrFour\ which reveal new correlated electron phenomena that are qualitatively different from  \SrTwo\ and  \SrThree\ due to the ferromagnetic order.
Our data are in general agreement with recent non-spin-polarized ARPES experiments \cite{Ngabonziza2020}, and add the new aspects of spin polarization, temperature dependence and greater resolution.
In addition, measurement of the momentum dependence of the spectra over a large range outside the Brillouin zone shows the nature of the bands in more detail.
Like has been shown before for \SrTwo\ \cite{Damascelli2000} and \SrThree\ \cite{Tamai2008},
we find that DFT calculations provide a general guide to the $k$-dependence of the bands,
which makes it possible to interpret the data and quantify the shifts and renormalization due to correlation not taken into account by the mean field DFT calculations.

The DFT calculations \cite{Gebreyesus2022} found very different occupancies of the various $t_{2g}$ bands.
There are wide electron-like minority-spin bands at the Fermi energy that are less than half-filled.
The majority spin bands are filled or almost filled with hole-like bands at the Fermi energy.
Spin-resolved ARPES shows definitively the DFT-predicted minority spin conduction bands, and almost filled majority spin bands that form a Fermi liquid with sharp hole-like Fermi surfaces at low temperature.
Moreover, there are  two narrow bands that are almost completely spin polarized with opposite polarization in different parts of the BZ,
minority spin near the zone center and majority spin near the zone boundary saddle point.
 Comparison to DFT reveals the narrow bands to originate from different tri-layer energy-split \dxzyz\ orbital character and also with different spatial localization in central and outer layers.


 The most striking discovery in the present work is the temperature dependence of the two  narrow bands  that reside just below the Fermi energy.  The narrow band peaks are very intense at low $T$ with $\sim$20 meV width, and dramatically decrease in amplitude up to and above $T_c$,  indicative of coherent spectral weight loss, while simultaneously their spin-polarization remains $>$60\% up to $T_c$.
The energies of the narrow bands are temperature-independent and there are only small temperature-dependent energy shifts of other spin-polarized bands, which is very different from a Stoner-like band picture and suggests localized moment behavior.
Such narrow bands near the saddle point have also been identified in \SrFour\ using STM-based quasiparticle-interference and high resolution spin-integrated ARPES \cite{Marques2024}. The results here provide addition information on the nature of the bands and for field-tuned vHS Lifshitz Fermi surface reconstruction for the in-plane metamagnetism proposed in Ref. \cite{Marques2024}.

The organization of this paper is to first present the low temperature spin-integrated and spin-resolved ARPES results, and then discuss the comparison to spin-polarized DFT calculations in order to identify the specific orbital origins of the key features.  Then temperature dependent ARPES results, also both spin-integrated and spin-resolved, are presented for the various narrow bands, Fermi-edge band crossings and oxygen bands.
The Discussion section then provides additional interpretation of the general question of itinerant versus localized magnetism, and the origins of the strong narrow band $T$-dependence 
in terms of coherent spectral weight loss and Hund metal spectral linewidth predictions,
before introducing the analysis of the layer-dependence based on DFT calculations
and experimental consistency with the photoemission structure factor.
Additional information is presented in the appendices.

\section{Methods} \label{methods}

 \subsection{Experimental}
Single crystals of \SrFour\ were grown using flux techniques described in Ref. \cite{Cao2007}.
Spin-integrated ARPES in the photon energy range of 30-150 eV was performed at the MERLIN beamline 4.0.3 of the Advanced Light Source (ALS) employing both linear horizontal (LH) and linear vertical (LV) polarizations from an elliptically polarized undulator. A Scienta R8000 electron spectrometer was used in combination with a six-axis helium cryostat goniometer in the temperature range of 10-150K with a total energy resolution of $\geq$15 meV and base pressure of $<$5$\times$10$^{-11}$ Torr. 
Variable temperature measurements were performed for warming with thermal equilibration at each step.

 Spin-resolved ARPES was performed at ALS beamline 10.0.1 using 56 eV and 76 eV LH-polarized x rays.
A Scienta-Omicron R4000 DA30 spectrometer with two Ferrum spin detectors \cite{Escher2011}
was used for data acquisition.
Samples were \textit{in situ} field-cooled from $>$120 to 14 K with $c$-axis field-alignment prior to cleavage.
Only $c$-axis spin-asymmetry was measured for this study.  More details about the spin-resolved and spin-integrated ARPES measurements are provided in Appendices \ref{appendix:spindet} and \ref{appendix:photon}.

 \subsection{Theory}
The ARPES results are compared to predictions of spin-polarized density functonal theory (DFT) calculations of the ferromagnetic
ground state of \SrFour\ using the pseudopotential, plane-wave implementation of DFT in Quantum Espresso  \citep{Giannozzi2009,Giannozzi2017} and using the PBEsol exchange-correlation functional \cite{Perdew2008}  without and with spin-orbit coupling as described in Ref. \cite{Gebreyesus2022}.
Additional LDA exchange-correlation functional predictions are provided in Appendix \ref{appendix:dft}.

\section{Low Temperature Electronic Structure}

\subsection{Spin-integrated ARPES} \label{spin-integrated}

\begin{figure*}[t]
\begin{center}
\includegraphics[width=18cm]{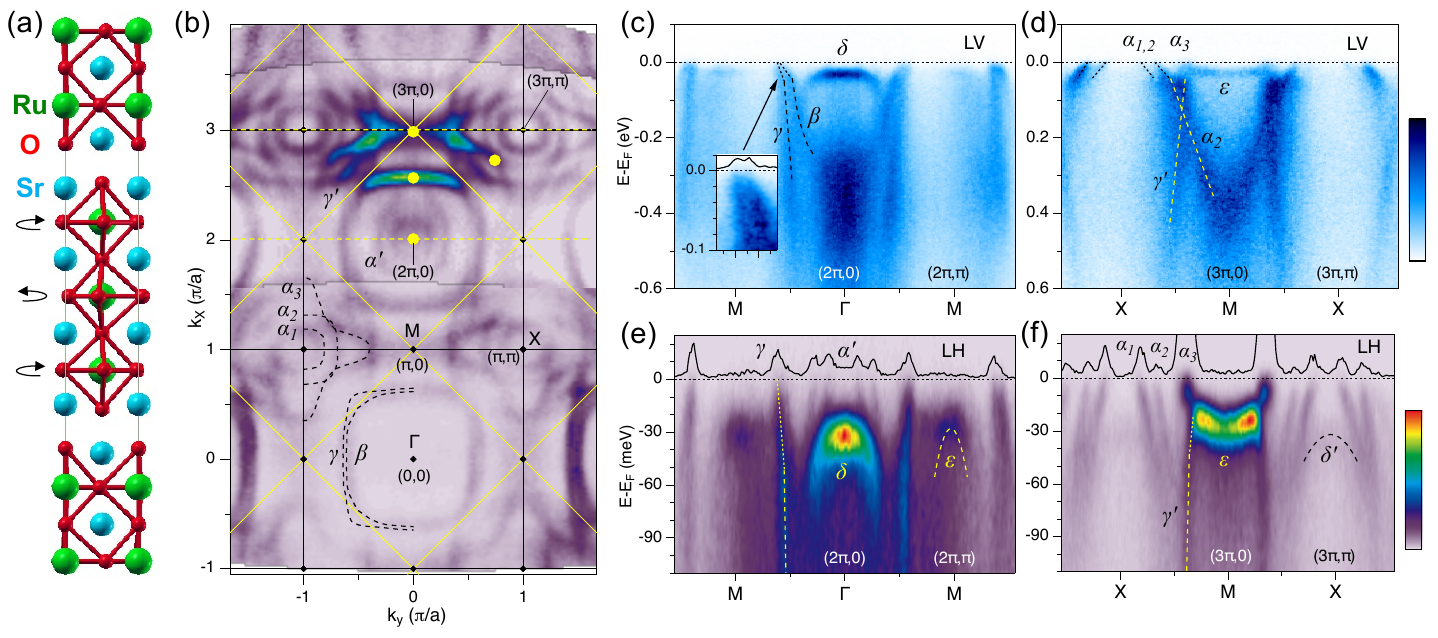}
\caption{
\textbf{ Spin-integrated ARPES electronic structure of low temperature \SrFour.}
(a) Side view of the three-layer crystal structure with oxygen octahedra rotation directions indicated.
(b) Fermi surface map using $h\nu=76$ eV linear-horizontal (LH) polarized x-rays.
(c,d) Wide band dispersion images using linear-vertical (LV) polarized x-rays for the two horizontal
dotted lines in (b) passing through (2\p,0) and (3\p,0).  
Inset in (c) shows a zoom of the two near-parallel \EF-crossing bands.
(e,f) Corresponding near-\EF\ band dispersion images using LH-polarization with Fermi-edge intensity profiles.
Dots in (b) indicate where spin-resolved ARPES spectra were measured.
Band and FS contour labeling follow \SrTwo\ BZ location and $d$-orbital origins for ($\alpha,\beta,\gamma$) and Ref. \cite{Ngabonziza2020} for ($\delta$).
Zone-folded bands are labeled with a prime ($^\prime$).
}
\label{fig_intro}
\end{center}
\end{figure*}

We first present low temperature constant energy high-symmetry
ARPES  maps and electronic band dispersions of \SrFour\ in Fig. \ref{fig_intro}.
The layered structure of this material exhibits opposite in-plane RuO$_6$ octahedral rotations between
the middle and outer layers, as illustrated in Fig. \ref{fig_intro}(a),
which results in an in-plane doubling of the unit cell and a zone-folding of the (unrotated) tetragonal electronic structure into a smaller 45\deg-rotated diamond-shaped Brillouin zone (BZ), shown as yellow lines in Fig. \ref{fig_intro}(b).
The spectral intensities of the zone-folded bands are observed to be weak in the photon energy range of these ARPES measurements.
Hence for this study, we use the tetragonal BZ labeling of \G\ for the (0,0) zone-center, M for the (\p,0) zone-edge, and X for the (\p,\p) zone-corner for discussion of experimental and theory results.
Note that for the rotated-BZ, X and M labels \cite{Gebreyesus2022} refer differently to (\p,0) and (\p/2,\p/2) $k$-points, respectively, and also M refers to (\p,\p) in the BZ for \SrOne\ \cite{Hahn2021}.


\subsubsection{Fermi Surfaces}

Figure \ref{fig_intro}(b) shows a spin-integrated Fermi-surface (FS) map spanning two BZs measured using a photon energy of 76 eV with  LH polarization,
chosen to maximize the signal as described in Appendix \ref{appendix:photon}.
A prominent feature in the FS map is the large zone-centered rounded-square
FS contour visible in both BZs, and especially enhanced in intensity along one edge in the second zone at $\sim$(2.5\p,0). This feature is actually composed of two wider-energy electron band dispersions with very different band minima, as shown in the band dispersion image in Fig. \ref{fig_intro}(c), using LV photon polarization.
The outer band ($\gamma$) has a light mass dispersion originating from below $-$0.7 eV, while the inner band ($\beta$) emerges from a high-intensity band minimum at $-$0.35 eV.
As discussed in the previous ARPES study \cite{Ngabonziza2020}, both bands have a heavier mass ``kink'' dispersion from $-$40 meV to \EF, resulting in a near-parallel Fermi velocity and small momentum-separation along $k_x$ and $k_y$ [inset in Fig. \ref{fig_intro}(c)] that becomes larger at the diagonal corners of the FS contour.

Another distinct feature of the low temperature band structure is a pair of concentric small square FS contours ($\alpha_{1,2}$)  located inside a flower-shaped FS contour ($\alpha_3$) at the tetragonal  X-point (\p,\p) zone corner, mostly clearly visible at (3\p,$\pm$\p) in Fig. \ref{fig_intro}(b). They originate from hole-bands, shown in Fig. \ref{fig_intro}(e) and \ref{fig_intro}(f), that are only sharp down to $-$40 meV and then become broader and less distinct down to $-$80 meV as they come closer to a strong intensity hole band ($\gamma'$) centered on M
that intercepts the shallow saddle-point electron band ($\varepsilon$).
The higher binding energy (BE) dispersion of the outer (\p,\p) hole band ($\alpha_{2}$), crossing the
hole band ($\gamma'$) and reaching $-$0.4 eV at M, is visible with LV polarization in Fig. \ref{fig_intro}(e).  Similar to the large  FS electron bands ($\gamma/\beta$),  the  (\p,\p) hole bands ($\alpha_{1,2}$) also have a distinct heavier mass Fermi velocity kink above $-$40 meV compared to its higher BE dispersion.

Table \ref{table1} summarizes the linear band velocities in units of eV-\AA.
 The \EF-crossing bands ($\gamma$, $\beta$, $\alpha_{1,2}$) exhibit kinks in their band dispersion in the energy range of 40$-$70 meV, and so high binding energy ($v$) and Fermi-velocity ($v_F$) linear slopes, below and above the kink energy, are evaluated.

{\setlength{\tabcolsep}{0.35em}
\begin{table}[b]
\caption{Low temperature band velocities in units of eV-\AA\ and effective masses in units of the free electron mass found by spin-integrated measurements for \SrFour. As discussed in the text, the two values of the velocities are for energies greater than (less then) the kink energy which is approximately 40 meV.  The theoretical values are from the DFT band structures discussed in Sec. \ref{DFT-comparison}. 
}
\begin{tabular}{ c cc cc c c l}
 \toprule
  band &  &  &  & $v$ & $v_F$  & DFT  & DFT  \\   
label & center & type & $m$* & \textit{ eV-\AA} & \textit{ eV-\AA}  & \textit{ eV-\AA}  & \textit{/expt}  \\
 \toprule
     $\delta$ & $\Gamma$(0,0) & h $\downarrow$ & 10 & -  & - & - & 1.7$\times$ \\
     $\varepsilon$ & M(\p,0)  & e $\uparrow$ & 15 & -  & - & - & 2.0$\times$ \\
        & M(0,\p)  & h $\uparrow$ & 2 & -  & - & - & 2.0$\times$ \\
     $\gamma$ & $\Gamma$(0,0) & e $\downarrow$  & - & 4.0 & 1.0  & 3.2 & 3.2$\times$ \\
     $\gamma^\prime$ & M(\p,0) & h $\downarrow$  & - & 4.0 & 1.0 & 3.2 & 3.2$\times$\\
     $\beta$ & $\Gamma$(0,0) & e $\downarrow$  & 2 & 1.4 & 0.5 & 2.5 & 5$\times$ \\
     $\alpha_1$ & X(\p,\p) & h $\uparrow$  & - & 1.0 & 0.55  & 1.7 & 3.1$\times$  \\
     $\alpha_2$ & X(\p,\p) & h $\uparrow$  & - & - & 0.42 & 1.8 & 4.3$\times$  \\
\botrule
\end{tabular}
\label{table1}
\end{table}
}

The octahedral rotation effect of zone-folding of tetragonal zone bands
is also visible in Fig. \ref{fig_intro}, from the presence of FS contours and bands labeled with a prime ($^\prime$).
The two concentric square FS contours ($\alpha$) at the (\p,\p) X-point appear with weak intensity  ($\alpha'$) at the (2\p,0) \G-point in Fig. \ref{fig_intro}(b),  and their sharp band dispersions are clearly visible at \G\ in Fig. \ref{fig_intro}(c) above the $-$30 meV hole band.
The converse zone-folding of the \G-point flat band ($\delta$) to X is also evident in a faint hole-like spectral intensity ($\delta$$^\prime$) at $-$30 meV  interior to the distinct hole bands at (3\p,$\pm$\p) Fig. \ref{fig_intro}(f).
Zone folding of the large rounded-square zone-centered FS contour ($\gamma$) is also observable ($\gamma'$) in Fig. \ref{fig_intro}(b), centered at all of the ($\pm$\p,$\pm$\p) and (3\p,$\pm$\p) points.
Similarly, the strong intensity light mass band {($\gamma'$) in Figs. \ref{fig_intro}(e) and \ref{fig_intro}(f) that
 is hole-like relative to M and intercepts the saddle-point band ($\varepsilon$),
is identified to be the zone-folded replica of the electron-like band ($\gamma$) centered on \G.

\subsubsection{Narrow bands just below the Fermi energy}

Two aspects of the spectra deserve special attention. As shown in Figs. \ref{fig_intro}(c)$-$\ref{fig_intro}(f), there are features approximately 30 meV below the Fermi energy at the zone center \G\ and the zone boundary saddle point M.  These bands are almost fully spin polarized as described later. 
The spin-integrated data in Figs. \ref{fig_intro}(c) and \ref{fig_intro}(e) reveal a shallow hole-band ($\delta$) at the \G-point  with $<$20 meV dispersion over $\sim$1/4 of the BZ along M-\G-M, corresponding to an effective mass of $\approx$10 meV as listed in Table \ref{table1}. 
This narrow band is discussed further in the following section where it is compared with bands from a spin-polarized DFT calculation, and its intrinsic linewidth discussed in more detail in Sec.
\ref{DiscussNB_Tdep}.

A second key high symmetry shallow band  exists at the (\p,0) M-point, where a -30 meV deep electron-like band ($\varepsilon$) shown in Fig. \ref{fig_intro}(d), that disperses $<$10 meV over $\sim$1/3 of the BZ along X-M-X.
This narrow band is observed to be strongly enhanced at the (3\p,0) boundary between second and third BZs for 76 eV and LH polarization as shown in Fig. \ref{fig_intro}(f).
Similar to \SrTwo, this tetragonal zone edge feature is actually a saddle-point van Hove singularity, with a much greater hole-like dispersion along \G-M as evident in Fig. \ref{fig_intro}(e).  
Thus, this band is very narrow along a line in $k$-space with highly anisotropic effective masses of $\sim$15 and $\sim$2 in the orthogonal directions.
The $T$-dependent hybridization of this vHS band ($\varepsilon$) with the zone-folded hole-like band ($\gamma'$) is detailed in Sec. \ref{TdepSP}.

\subsection{Spin-resolved ARPES} \label{spin-resolved}

We now turn attention to identifying the spin polarization of different key ARPES features identified by yellow dots in Fig. \ref{fig_intro}(b).
The representative spin-resolved ARPES spectra in Fig. \ref{fig_spin4}, for the $c$-axis cleaved
crystal, show only the $c$-axis vector component of the spin polarization
parallel to the field-cooled magnetization direction along the magnetic easy axis.
For the \G-point flat band spectrum at the (2\p,0) hotspot ($\delta$) measured with a photon energy of 56 eV,
the narrow $-$30 meV peak shows a strongly spin-minority polarization (i.e. spin direction opposite to the global magnetic moment) in Fig. \ref{fig_spin4}(a).
At higher binding energy, the spectrum shows a sign reversal to spin-majority polarization with uniform 20\% amplitude  across a broad spectral hump centered at $-$0.3 eV.
In contrast, the narrow peak ($\varepsilon$) at the (3\p,0) M-point, measured at 76 eV,  shows a strong spin-majority polarization in Fig. \ref{fig_spin4}(b), opposite to that of the \G-point narrow band, but with a same-sign $\sim$30\% spin-majority polarization at higher binding energy.
While the net minority spin polarization at the narrow band maximum amplitudes are evaluated to be only 50\%-60\%, isolation of the spin-dependent peak amplitudes from modeled background intensities (dashed lines), results in higher spin polarizations (see details in the Appendix \ref{appendix:spindet}). 

Spin-resolved spectra were also acquired at two other larger band-velocity \EF-crossing $k$-points.
The spectra at the large zone-centered rounded-square FS contour ($\gamma/\beta$), measured at $\approx$(2.5\p,0)
and 76 eV, shows a distinct spin-minority composition in Fig. \ref{fig_spin4}(c)
similar to that of the \G-point flat band,
and eventual sign-reversal to spin-majority polarization below $-$0.5 eV.
The spectra at the intermediate-sized square FS contour ($\alpha_2$) close to the (3\p,\p) X-point, measured at 56 eV in Fig. \ref{fig_spin4}(d), shows a spin-majority polarization similar that of to the nearby M-point narrow band, but with distinctly weaker spin polarization.

Assuming same-sign spin polarization along FS contours and ignoring weak intensity zone-folded bands, a pattern of near-\EF\ states emerges of spin-minority polarization close to the zone-center and spin-majority polarization along the tetragonal zone boundary.
In the next section, we will see that this is a general consequence of having bands dispersing upward from the zone center with an exchange splitting nearly as large as the bandwidth.
A uniform spin-majority polarization at high binding energies is also consistently observed. Its origin is discussed later.

\begin{figure}[ht]
\begin{center}
\includegraphics[width=9cm]{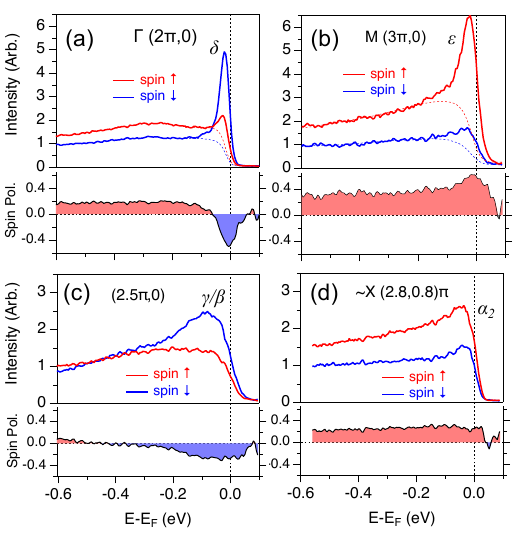}
\caption{
\textbf{Spin-resolved ARPES at low temperature.}
Spin-up and spin-down spectra measured at 14 K and the net spin polarization (Spin Pol.) at four selected $k$ locations: (a) the (2\p,0) \G-point,
(b) the (3\p,0) M-point saddle-band,
(c) the large FS contour on \G,
and (d) a square hole FS contour
centered on the (3\p,\p) X-point.
The four $k$ points are identified as yellow dots in Fig. \ref{fig_intro}.
All measurements use LH polarization excitation and are overall scaled to unity at high binding energy for the weaker intensity spin-minority background.
Dashed lines show background profiles used for additional quasiparticle peak analysis.
}
\label{fig_spin4}
\end{center}
\end{figure}

\section{Comparison to DFT} \label{DFT-comparison}

Together with recent spin-polarized DFT calculations \cite{Gebreyesus2022}, the present experimental results allow us to understand the major features of the electronic states.
Like the non-magnetic ruthenates \SrTwo\ and \SrThree, which have been studied extensively \cite{Tamai2019,Allan2013},
the bands near the Fermi energy are derived mainly from the Ru d $t_{2g}$, \dxy\ and \dxzyz\ manifold of states mixed with oxygen p states, but the results are very different due to the tri-layer structure and the magnetism.
In this section we compare with the DFT calculations using the PBEsol functional reported in Ref.\cite{Gebreyesus2022}.  Similarities and differences using other functionals are discussed in Appendix \ref{appendix:dft}.

In all cases, the dispersion of the bands in the plane starts at the zone center (0,0) and rise steeply to a maximum at (\p,\p) with a saddle-point at (\p,0) which leads to a large density of states (DOS).
The \dxy\ bands disperse in both directions, and the \dxzyz\ bands have one dimensional character, each dispersing strongly in one direction and nearly flat in the other.
In \SrFour\ the three planes of Ru and oxygen atoms are strongly coupled to form bonding (B), nonbonding (NB) and antibonding (AB) \dxzyz\ bands. This is analogous to the bonding and antibonding bands of the two-layer material \SrThree;
but there is a important difference because the central layer is not equivalent to the outer two layers, whereas the two layers are equivalent in \SrThree.

The rotations of the oxygen octahedra double the unit cell leading to the smaller Brillouin zone as shown in Fig. \ref{fig_intro}, the same as in \SrThree, which also has rotated octahedra and a cell that is doubled in the same way.  However, as illustrated in Fig. \ref{fig_intro},  there are only weak effects in the observed intensities and the data can be analyzed in an extended range of $k$-space corresponding to the larger tetragonal Brillouin zone, with weak replicas due to the doubling of the cell. This is indicated in Fig. \ref{fig_spindft}(a) where the bands from \cite{Gebreyesus2022} are ``unfolded'' schematically to show the regions of $k$-space where the intensity is large with  weak replicas indicated by light lines.

The qualitative difference in \SrFour\ is the ferromagnetism which leads to splitting of the bands into minority- and majority-spin bands.  Since there is a large moment, the majority-spin bands have large occupation so that the Fermi energy is near the top of the bands.
In contrast, the minority-spin bands have reduced occupation and the Fermi energy is in the lower part of the bands.
As shown in Fig. \ref{fig_spindft}(a), the wide electron-like minority-spin bands disperse steeply upward from (0,0) and  cross the Fermi energy.
These can be identified with the experimental bands that disperse upward in Fig. 1(d) and form the square-shaped Fermi surface that fills most of the BZ in Fig. 1(b), and are found to be minority spin in Fig. 2(c).
This is a robust result of the theory that depends only on characteristic shapes of the \dxy\ and \dxzyz\ bands, and the ferromagnetic order.

\begin{figure*}[ht]
\begin{center}
\includegraphics[width=18cm]{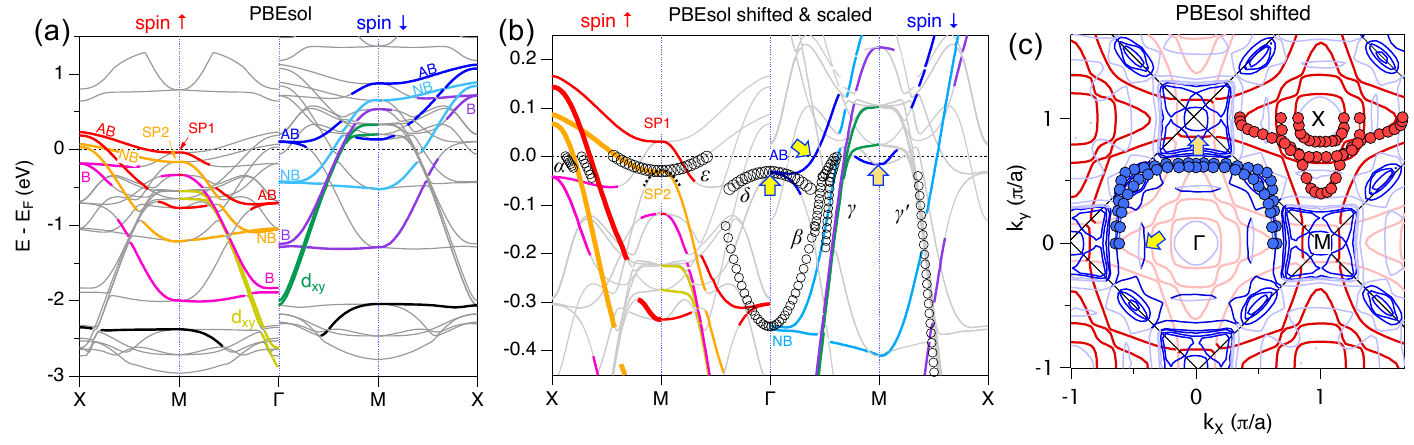}
\caption{
\textbf{Spin-resolved DFT and comparison to ARPES.}
(a) DFT GGA-PBEsol theory calculation of spin-up/down band structure with $\approx$0.9 eV exchange splitting, resulting in the AB band at \G\ just above \EF\ and the
saddle-point band SP1 crossing below \EF\ at M.
(b) Energy-shifted and scaled PBEsol bands with $\approx$0.6 eV exchange splitting, resulting in the AB band  at \G\ below \EF, and the SP2 band crossing below \EF\ at M.
Energy scale renormalization by $\approx$2$\times$ then provides good agreement to experimental bands (open circles).
(c) Comparison of experimental FS contours (filled circles) to PBEsol contours including the energy shifts in (b).
Solid colored lines emphasize the tetragonal band or contour origin, while lighter gray lines correspond to features with octahedral-rotation and zone-folding origins.
Arrows in (b) and (c) indicate two key disagreements that are resolved with the inclusion of spin-orbit coupling (see Appendix \ref{appendix:SOC}).  
}
\label{fig_spindft}
\end{center}
\end{figure*}

The prediction that the majority-spin bands are full or nearly full is also a robust result of the DFT calculations, and since the effects of band folding are small, we expect strong intensity in ARPES measurements  in the X-point region around (\p,\p). This provides an interpretation for the majority-spin bands observed in the experiment; however, the degree of filling depends on the energy difference between the majority and minority-spin bands and effects of correlation. As discussed below, present-day theory is not at the point where the energies can be predicted with accuracy of a few tenths of an eV, and we will use the experimental measurements as a guide.

There are two other definite predictions of the theory that provide the basis for understanding the spin-polarized narrow bands near the Fermi energy around \G\ and M
which are observed in the experiments.  The first point is that the only states with large intensity near (0,0) are spin-minority bands; as indicated in Fig. \ref{fig_spindft}(a), the spin-majority bands with large intensity near \G\ are well below the Fermi energy and any majority spin bands are due to the weak effects of folding.
Thus, even though there are several minority-spin bands in this region, with uncertainties in the theory and the fact that energies are affected by correlation beyond the DFT calculations, nevertheless, it is clear that the narrow peak observed in the experiment should be derived from bands that are overwhelmingly minority spin, as observed in the experiment.  The second conclusion from the calculations relates to the states near the saddle-band M-point.
As is clear from Fig. \ref{fig_spindft}(a), the theoretical calculation finds majority spin bands in this region below the Fermi energy, the same bands that form the majority-spin Fermi surfaces around X.
We can expect that the narrow peak observed just below the Fermi energy observed in the experiments is formed primarily from these bands, in agreement with the majority spin-polarization found in the experimental results.

A more quantitative analysis of the experimental results provides more information that helps to narrow down the uncertainties in the theory, and we have analyzed the comparison with the calculations in more detail. Comparison with the DFT calculations provides a consistent interpretation of the experimental results for the narrow bands near \G\ and M and the Fermi surfaces around X,
as long as we allow for small shifts of the calculated bands and renormalization of DFT band energies close to \EF, similar to the other Sr ruthenates.
 Figure \ref{fig_spindft}(b) shows the bands near the Fermi energy, with ARPES
bands (open circles) compared to the PBEsol calculation with $-$0.16 eV (+0.10 eV) energy-shift corrections to the minority (majority) spin bands, and a 2$\times$ energy
scale renormalization which then provides good agreement to the heavy effective mass dispersions of both the \G-hole and M-electron narrow bands as shown in Table \ref{table1}.
The effect of the downward shift of the energy on the minority-spin bands is to move the band at \G\ from just above \EF\ in the DFT calculation shown in Fig. Fig. \ref{fig_spindft}(a) to just below as shown in Fig. \ref{fig_spindft}(b); together with the renormalization, this provides an explanation for the narrow band $(\delta)$ at \G\ shown in Figs. \ref{fig_intro}(c), \ref{fig_intro}(d), and \ref{fig_spin4}(a), and represented by the line of circles near \G\
in Fig. \ref{fig_spindft}(b).
The effect of the upward shift of the majority-spin bands is to move the AB state at M
labeled SP1 in Fig. \ref{fig_spindft}(a) to above \EF\ and the state labeled SP2 to just below \EF, where it provides an explanation for the observed majority-spin band  at $-$30 meV ($\varepsilon$) in Figs. \ref{fig_intro}(e) and \ref{fig_spin4}(b). The shift leads to better agreement with experiment, which finds only one band below \EF\ at M
in this energy range, and it also improves the agreement for the majority-spin bands (shown as the lines of circles) that cross the Fermi energy  along M-X.
The Fermi surfaces for the shifted bands are shown in Fig. \ref{fig_spindft}(c), which shows good overall agreement with the major features of the experimental results.

However, the shifts of the bands do lead to two major
discrepancies with experiment, highlighted by arrows in Fig. \ref{fig_spindft}(b) and \ref{fig_spindft}(c).
First, the PBEsol calculations predict a two-fold band degeneracy just below \EF\ at \G,
which results in the prediction of both downwards and upwards dispersing AB branches, and an electron-like \EF\ crossing which is not observed experimentally.
Second, the downward shift leads to spin-minority bands at \EF\ in a region around M
for most combinations of energy-shift and energy-scaling corrections.
These effects lead to bands and pieces of Fermi surface marked by arrows in Figs. \ref{fig_spindft}(b) and \ref{fig_spindft}(c), which are not observed in the experiments.

The inclusion of spin-orbit coupling provides a resolution of these discrepancies, as described in detail in Appendix \ref{appendix:SOC}.
Since the spin-orbit coupling mixes the spins (quantitatively $<$5\%),  it is not a simple matter to identify the majority and minority spin projections and to manually shift the bands independently, as performed in Fig. \ref{fig_spindft}(b).
The SOC effects are not large overall, but there are major consequences for states near the Fermi energy. One is that the degeneracy of the AB states is lifted so that the upward curving band at \G\ is above the Fermi energy, consistent with experiment which did not observe such a band.
Additionaly, the energy of the minority-spin bands at M are raised
to above the Fermi energy.  The resulting Fermi surfaces shown in  Appendix \ref{appendix:SOC}
are in quantitatively better agreement with experiment, with the absence of the FS contours marked by arrows in Fig. \ref{fig_spindft}(c).

The shifts of the majority-spin bands upward and the minority-spin downward relative to the Fermi energy has the effect of reducing the differences between both the energies and the occupations of the majority and minority bands.  The average energy difference, called the exchange energy, is a consequence of interactions which are taken into account in approximate ways depending on the functional used in the calculation.
The shifts of the bands calculated using the PBEsol functional amount to a reduction of the exchange splitting from $\sim$0.9 eV to about $\sim$0.6 eV.
This results in a smaller spin moment and provides an
explanation for the discrepancy between PBEsol prediction of a 1.77 $\mu_B$ average moment per Ru site \cite{Gebreyesus2022} and the significantly smaller experimental value of 1.1 $\mu_B$/Ru \cite{Cao2003}.

A spin-polarized DFT calculation using the LDA exchange-correlation functional
finds a reduced average moment of 1.36 $\mu_B$ and a corresponding reduced exchange splitting of $\sim$0.6 eV
which is a closer match to the experimental ARPES.
 A comparison of DFT predictions using LDA, PBEsol, and PBEsol+$U$ functionals is provided in Appendix \ref{appendix:dft}.
 The better agreement of the LDA functional to experiment in \SrFour, the systematic trend towards increasing exchange splitting and larger moments for LDA $\rightarrow$ PBEsol $\rightarrow$ PBE96 $\rightarrow$ DFT+$U$, and the prediction of a large moment half-metal groundstate (not observed) for small $U$ is consistent with previous comparative studies of DFT exchange-correlation functionals for magnetism in \SrOne\ \cite{Granas2014}, \SrThree\ \cite{Rivero2017} and ferromagnetic transition metals \cite{FuSingh2019}.

\section{Temperature dependence} \label{Tdep}

\subsection{Narrow band $T$-dependence and \G\ spin-polarization} \label{TdepNB}

\begin{figure*}[t]
\begin{center}
\includegraphics[width=18cm]{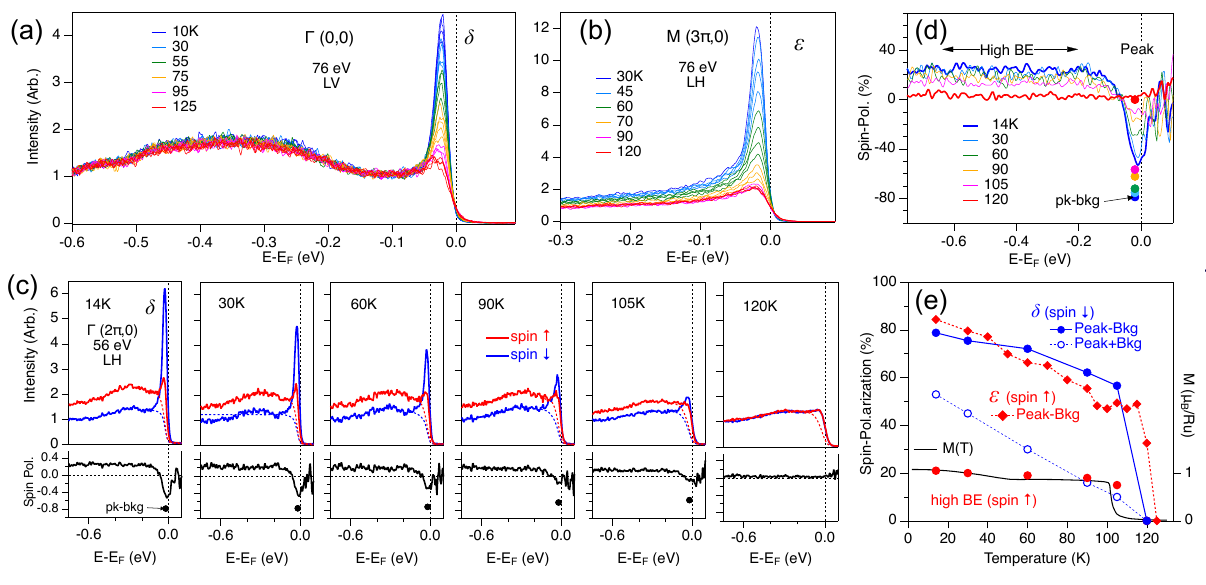}
\caption{
\textbf{Narrow band temperature dependence and spin polarization.}
Spin-integrated spectra at (a) the (0,0) \G-point and (b) the (3\p,0) M-point saddle-band exhibiting dramatic temperature dependent amplitude reductions of the strong narrow peaks warming to above $T_c$.
(c) Spin-polarized (2\p,0) \G-point spectra and spin-polarization profiles for selected temperatures from 14$-$120 K.
(d) Comparison of the spin-polarization profiles from (c).
(e) Summary of the spin polarization of the average of the high binding energy region shown in (d), the narrow peak amplitude, and the background-subtraction corrected spin polarization corresponding to the dots in (c) and (d), including an additional 1.11$\times$ geometry-correction.
A $c$-axis bulk magnetization profile \SrFour\ is provided for reference.
Spin-polarization analysis of the $T$-dependent spin-majority (3\p,0) M-point is also plotted in (e) with a 1.21$\times$ geometry-correction.
}
\label{fig_tdep}
\end{center}
\end{figure*}

We next turn our attention to the temperature dependent amplitude and spin-polarization of the prominent narrow bands at $-$30 meV.  Figures \ref{fig_tdep}(a) and \ref{fig_tdep}(b) present the fine temperature-step spin-integrated evolutions of the narrow band amplitudes.
Both narrow bands exhibit dramatic amplitude reductions towards higher temperature and $T_c$,
with an approximately linear dependence. 
However, in the course of the line-shape fitting described in Sec. \ref{DiscussNB_Tdep}, it was found that the approximate linearity is not intrinsic, but arises from the effects of our experimental resolution masking a low $T$ nonlinear dependence.  Therefore we give it no special significance.

The spin-resolved temperature evolution of the \G\ point was measured at the 56 eV (2\p,0) $k$ location with spectra presented in Fig. \ref{fig_tdep}(c) for $\approx$30 K steps from 14 to 120 K, along with their spin-polarization energy profiles, which are overplotted for comparison in Fig. \ref{fig_tdep}(d).
The strong spin-minority narrow band amplitude is observed to monotonically decrease with temperature, consistent with the spin-integrated measurement, whereupon, as expected, the spin polarization of both the peak and high BE background go to zero above $T_c$.
The high BE spin polarization, averaged over the energy interval of $-$0.2 to $-$0.6 eV,
exhibits a weak monotonic decline from 21\% to 15\% at $T_c$ as summarized in Fig. \ref{fig_tdep}(e), along with comparison to a bulk magnetization curve \cite{Zheng2018} that exhibits a similar weak decline.
In contrast, the raw spin-polarization analysis of the narrow peak exhibits a steady monotonic decline, seemingly correlated to that of the amplitude variation.
However, with consideration of a background subtraction to isolate the spin-asymmetry of the narrow peak, and also with a second BZ large angle 1/cos$\theta$ correction factor of 1.11
 (see Appendix \ref{appendix:spindet} 
 for more details), a modified $T$-dependent spin-polarization profile of the narrow peak in Fig. \ref{fig_tdep}(e) also exhibits a weak decline from 80\% at low $T$ to 60\% close to $T_c$,
 that scales very well with that of the high BE spin-polarization profile.

\subsection{Saddle-point polarization and hybridization}\label{TdepSP} 

\begin{figure*}[t]
\begin{center}
\includegraphics[width=18cm]{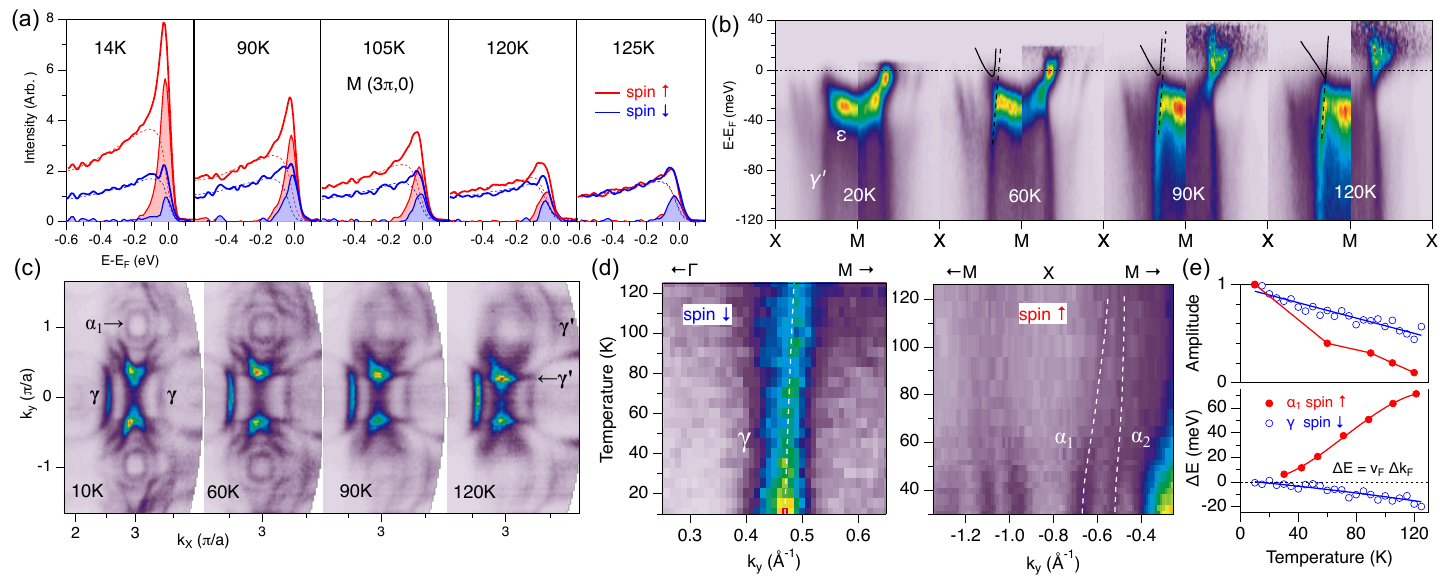}
\caption{
\textbf{Zone boundary temperature dependence and saddle-point spin-polarization.}
(a) Selected spin-polarized spectra at M (3\p,0) showing the reduction to zero \textit{above} the bulk $T_c=105$ K.
(b) Spin-integrated 76 eV X-M-X spectra for selected temperatures, with intensity normalization along energy (right side), showing the evolution of hybridization between a zone-folded spin-minority hole-like band ($\gamma '$) that crosses the spin-majority saddle-point band ($\varepsilon$) near (3\p,0).
(c) $T$-dependent Fermi surface maps along the X-M-X zone boundary showing the strong decoherence evolution of the spin-majority contours relative to the uniformly sharp spin-minority contours.
(d) Comparison of the $T$-dependent Fermi-edge momentum intensity profiles 
for the spin-minority $\gamma$ electron band and the spin-majority (\p,\p) X-point hole bands ($\alpha_{1,2}$). Dashed lines highlight varying \kF.
(e) Summary of the \kF\ amplitudes and \kF\ shifts converted to energy shifts using Fermi velocities from Table \ref{table1}.
}
\label{fig_tdepsp}
\end{center}
\end{figure*}

 The spin-resolved temperature evolution of the spin-majority (3\p,0) M-point saddle-band at 76 eV was also measured, with finer temperature steps.   Selected spin-resolved spectra at M are shown in Fig. \ref{fig_tdepsp}(a), along with the background profiles and narrow band peaks whose areas are used to generate the spin-polarization data points plotted in Fig. \ref{fig_tdep}(e). It shows a similar $>$80\% spin polarization at low temperature and also confirms that the spin-polarization goes to zero a full 20K higher than the bulk $T_c$.

 As discussed in Fig. \ref{fig_intro}(e) the low temperature X-M-X zone-boundary cut at (3\p,0) is dominated by spin-majority hole bands ($\alpha_{1,2}$) crossing \EF\ at X and the strong $-$30 meV saddle-band van Hove singularity ($\varepsilon$) at M.
 In addition, a strong intensity hole-like band,
 identified as a zone-folded spin-minority \dxy\ band ($\gamma^\prime$), is observed to disperse up to the spin-majority ($\varepsilon$) band, with no clear extrapolation of the band to \EF.  The strong intensity at the intersection point,  and also an abnormal ``horned''-shaped vertical dispersion right at \EF, suggests a strong hybridization between the two bands.

 In Fig. \ref{fig_tdepsp}(b),  we visualize the temperature dependence of this near-\EF\ hybridization region enhanced by normalization of the intensity sum at each energy (right side at each temperature).  With increasing temperature, we observed within our energy and momentum resolution, that the `horned'-shaped vertical dispersion bifurcates into two bands above \EF, with its band minimum close to \EF.  The outer edge of these two tiny electron-like pockets follow the parabolic electron dispersion of the ($\varepsilon$) saddle-band, while the inner edge has the band velocity of the high BE spin-minority hole-band, but displaced to larger Fermi momentum \kF.  At the highest temperature of 120 K, where the spin-majority saddle-band intensity has dramatically decreased, the spin-minority band appears to disperse linearly up through \EF, unaffected by the band crossing.  This suggests that the decoherence of the spin-majority band reduces the hybridization with the still highly coherent spin-minority band.

 The effects of the reduced hybridization on the dispersion of the zone-folded spin-minority band is also evident in the corresponding temperature-dependent FS maps in Fig. \ref{fig_tdepsp}(c).  Two nearly straight horizontal contours  ($\gamma '$), obscured at low temperature by the strong intensity tips of the ($\alpha_3$) flower contour, emerge more distinctly at high temperature, forming a fourfold symmetric square centered on M, comprised of the edges of the original and zone-folded \G-centered electron bands.
The low temperature hybridization gap in this zone boundary region of the M-point has recently been fully visualized with lower photon energy high-resolution ARPES \cite{Marques2024} as discussed later.

\subsection{Spin-dependent \EF-crossing $T$-dependence} \label{TdepPP}

Next we focus on isolated Fermi-edge crossing bands without the complexity of the previous zone-folded band-crossing hybridization. In the series of Fermi-surface intensity maps in Fig. \ref{fig_tdepsp}(c), a key visual observation is how the X-point zone corner spin-majority hole band FS contours ($\alpha_{1,2}$) become dramatically broader at high temperature, especially in contrast to the spin-minority $\gamma$ FS contour, which appears to remain strong and sharp up to high temperature.  
 To visualize and quantify this spin-dependent behavior, we plot finer temperature step Fermi-edge momentum profile cuts along $\Gamma$-M and along M-X-M in Fig. \ref{fig_tdepsp}(d), extracted from the same data sets as Figs. \ref{fig_tdep}(a) and \ref{fig_tdep}(b), respectively.
Here we discover that, in addition to the clear more rapid decoherence of the spin-majority hole bands, the \EF-crossings exhibit distinct \kF\ shifts with increasing temperature.   
The \kF\ shifts correspond to increasing FS size in both cases, and 
can be rationalized as resulting from an upwards energy shift of the spin-majority hole bands, and a downwards energy shift of the spin-minority electron band, which in turn is suggestive of a weakening of the Stoner exchange splitting. 
The corresponding $T$-dependent energy shifts, estimated by $\Delta E$$\approx$\vF $\Delta$\kF\ using the Fermi velocities of the bands from Table \ref{table1}, are plotted in Fig. \ref{fig_tdepsp}(e) along with a quantification of the \kF\ intensity amplitudes reflective of the coherency reduction. 
In addition to the large difference in coherency $T$-evolutions for the two spins,  there is a large asymmetry in the magnitude of the energy shifts of $>$60 meV of the spin-majority bands compared to $<$20 meV shift of the spin-minority band up to $T_c$.  Both $T$-dependent energy shifts, though, are far less than the low $T$ Fermi-edge exchange splitting of $\approx$0.6 eV determined in Fig. \ref{fig_spindft}.  Origins of spin-asymmetric $T$-dependencies are further discussed later.
 \\

\section{Oxygen-band spin-polarization and $T$-dependence} \label{TdepO}

\begin{figure*}[t]
\begin{center}
\includegraphics[width=18cm]{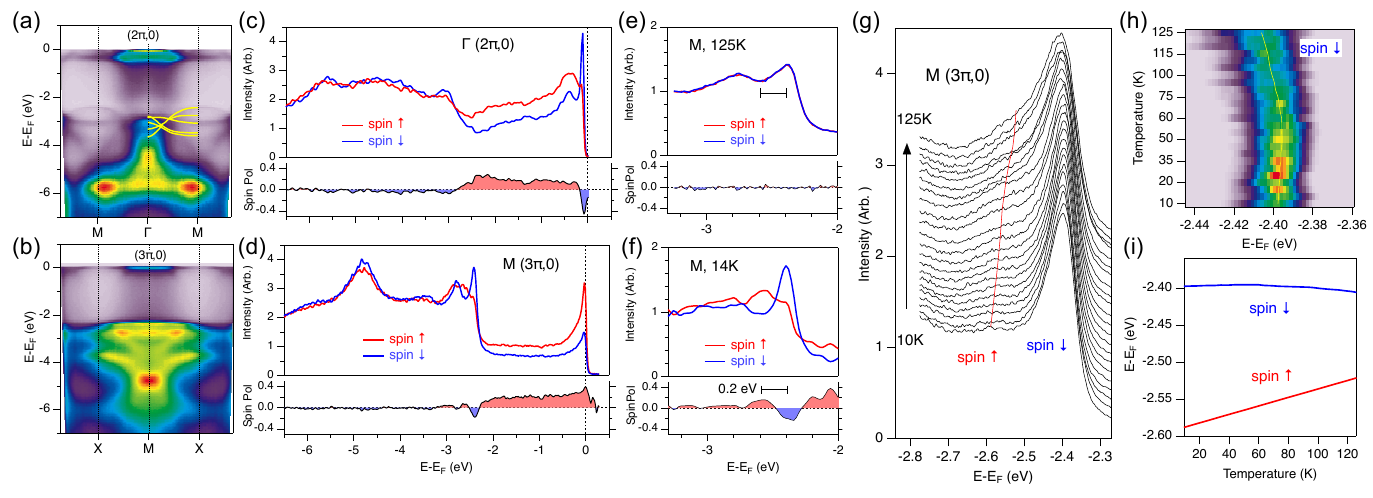}
\caption{
\textbf{Oxygen band spin polarization and temperature-dependent exchange splitting.}
(a,b) Wide-energy spin-integrated ARPES at (a) \G\ (2\p,0) and (b) M (3\p,0) highlighting the sharp bands at the top of the oxygen manifold, with agreement to DFT bands (with 0.7$\times$ scaling and a $-$0.26 eV shift).
(c,d) Wide-energy spin-ARPES at (c) \G\ and (d) M at 76 eV highlighting the extent of the spin-majority-polarized Ru-$d$ band incoherent background down to the top of the O-bands where the sharp bands show spin-minority polarization.
(e,f) Spin-ARPES spectra at the top of the M-point oxygen flat band at (e) 125 K and (f) at 14 K showing evidence for a $-$0.2 eV energy spin-majority peak.
(g) Spin-integrated $T$-dependent ARPES of the M-point oxygen bands showing the evolution of the high binding energy (spin-majority) shoulder.
(h) Intensity plot of the sharp spin-minority O band showing showing a very small energy shift.
(i) Summary of the asymmetric $T$-dependent energy shifts of the spin-majority and spin-minority O bands.
}
\label{fig_tdepo}
\end{center}
\end{figure*}

Significant spin polarization of oxygen bands is predicted by DFT to occur via strong hybridization of oxygen states with Ru $t_{2g}$ states in the 2 eV valence region below \EF.
In addition, model analysis of neutron diffraction measurements predict a 0.2 $\mu_B$ O-moment contribution to the total moment per Ru site, and with a 3$\times$ larger moment for basal (Ru-layer) O than apical O sites \cite{Forte2019}.   This can be understood from the large wavefunction overlap of basal O $p_{x,y}$ orbitals with the Ru \dxy\ orbitals, which have large dispersion from the top of the O-band manifold to \EF.

Here we present the  novel discovery of distinct spin-polarized narrow oxygen bands of nearly pure O-character at $\sim$2.5 eV binding energy at the M-point in \SrFour, and characterization of their $T$-dependent exchange splitting.
The wide-energy spin-integrated ARPES cut along M-\G-M at (2\p,0) in Fig. \ref{fig_tdepo}(a) reveals sharp `eye'-like band dispersions at the top of the O manifold at the M-point.
The distinctive dispersion is reproduced by the DFT band calculations with small quantitative scaling and energy shifts.
An orthogonal wide-energy ARPES cut along X-M-X at (3\p,0) in Fig. \ref{fig_tdepo}(b) shows the top of this band to be very flat along the zone boundary.
The flat O-band dispersion along  M-X is also reproduced in the spin-polarized DFT plot in Fig. \ref{fig_spindft}(a).
Furthermore, DFT orbital character projections reveal that this narrow M-point band is composed of basal O $p_z$ orbitals and apical O $p_{x,y}$ orbitals,  both of which have minimized wave function spatial overlap with the Ru \dxy\ orbitals.
The result is a very weak Ru-$d$ character mixing of this narrow O band at M, in sharp contrast to the strong mixing of basal O $p_{x,y}$-Ru \dxy orbitals discussed above.

 The wide-energy spin-resolved ARPES spectra, including the entire O-band manifold, in Fig. \ref{fig_tdepo}(c) and (d) highlight the extent of the spin-majority-polarized Ru-$d$ band incoherent background down to the top of the O bands where a reversal of the sign of the spin polarization is observed both at (2\p,0) and (3\p,0).
 In particular, the sharp bands at M have distinctly spin-minority polarization.
 A zoom in on this region in Fig. \ref{fig_tdepo}(f) shows evidence for a spin-majority peak in the valley between the two spin-minority peaks, which suggests a 0.2 eV exchange splitting, somewhat smaller than the DFT (PBEsol) prediction of 0.33 eV spin-splitting in Fig. \ref{fig_spindft}(a).
 Figure \ref{fig_tdepo}(f) shows how the net spin polarization is gone at 120K and with a different asymmetric  lineshape of the leading edge peak.

 We investigate this $T$-dependent line-shape evolution with higher resolution spin-integrated measurements in Fig. \ref{fig_tdepo}(g).  The much weaker spin-majority peak is observed to increase in intensity and shift about 50 meV to lower binding energy towards the strong spin-majority peak which has at most only a 5 meV energy centroid shift, illustrated in the intensity image in Fig. \ref{fig_tdepo}(h).  A summary plot of these $T$-dependent peak shifts is given in Fig. \ref{fig_tdepo}(i). The spin-dependent magnitudes of the O-band energy shifts are similar to that of the \EF-crossing bands analyzed in Fig. \ref{fig_tdepsp}(e).

\section{Discussion} \label{discussion}

 \subsection{Itinerant versus localized magnetism} \label{DiscussDuality}


The traditional simplified pictures are the Stoner model of itinerant band ferromagnetism and, at the other extreme, localized spins which are ordered in the ferromagnetic phase.
In the Stoner picture, there are well-defined bands at all temperatures, and the energy difference between the bands for the two spin states, called the exchange energy, decreases with temperature until the bands become degenerate above the transition temperature $T_c$.
On the other hand, the picture of localized spins is that they persist at all temperatures and become disordered above $T_c$. The energies of the states remain the same even if the directions of the moments are disordered with no preferred direction. In addition, one expects localized states to lead to broadening of band-like features, so that the spectra would not show sharp features with dispersion as a function of momentum.
In a Fermi-liquid metal, however, there is another important consideration. No matter what is the nature of the states, at low temperature there should be well-defined bands near the Fermi energy and a sharp Fermi surface satisfying the Luttinger theorem, i.e., even states that act as localized moments at high temperature form coherent delocalized states at low temperature.   Additionally, for the two ferromagnetic ruthenates \SrOne\ and \SrFour, the magnetization at low temperature should be determined by the difference in the volumes of the Fermi surfaces for the two spin states multiplied by $\mu_B$.

The experimental data presented here for \SrFour\ in Figs. \ref{fig_intro}  and \ref{fig_spin4} at low temperature shows clearly Fermi surfaces for both spins, with mainly filled majority and mainly empty minority spin bands.
To understand better the nature of the magnetic moments we must turn to measurements as a function of energy (away from \EF) at low temperature  and/or the behavior as a function of temperature.
As is also true for \SrOne \cite{Shai2013,Hahn2021}, away from \EF\ the spectra at low $T$ in the Ru-$d$ valence band region down to $-$2 eV is broad and incoherent.  Both the spin majority and spin-minority narrow band peaks shown in Fig. \ref{fig_tdep} have $T$-independent energies, suggesting a Heisenberg-like $T$-independent exchange splitting for both.  The spectra show no features of split-off Hubbard bands or atomic multiplets, which is consistent with the expectation for 4$d$ states for which the interactions are not as large as those for 3$d$ states. The $T$ dependencies are discussed in the following section where it is concluded that on the whole the results are consistent with local moments and the Hund metal picture.

However, the spin-majority hole bands ($\alpha_{1,2}$) and the spin-minority $\gamma$ band that cross \EF,  do have momentum and energy shifts with increasing $T$, as shown in Figs. \ref{fig_tdepsp}(d)$-$\ref{fig_tdepsp}(f), signaling some decrease of exchange splitting, although well short of a full $\sim$0.6 eV Stoner collapse at $T_c$.  Similarly, the spin-polarized oxygen flat band at M 
in Fig. \ref{fig_tdepo} with 0.2 eV exchange splitting, shows a similar incomplete 50 meV energy shift of the spin-majority peak, and much smaller shift of the spin-minority peak, also favoring a localized description of the magnetism.

A recent spin-polarized ARPES study of study of epitaxial films of \SrOne\ 15 unit-cell thick reported a peak at \G\ just below the Fermi energy quantitatively similar to the narrow band found here,  
but with a completely different zone-folding origin.
The spin polarization was found to be strongly dependent on momentum around the Fermi level with states whose energies shift with temperature,
whereas there is less $T$-dependence at higher binding energies \cite{Hahn2021}.  Based on analysis of such signatures, a ``dual ferromagnetism'' model was proposed, in which the moments of the spin-minority electrons are itinerant and the moments of the spin-majority electrons are localized.

Our results for \SrFour\ also manifest a kind of duality with polarization that is momentum dependent near the Fermi energy, by virtue of narrow bands having opposite polarization at the \G\ and M points; however, both the spin majority and spin-minority narrow band peaks shown in Fig. \ref{fig_tdep} have $T$-independent energies, suggesting a Heisenberg-like $T$-independent exchange splitting for both.
This suggests quite a different sense of ``duality'' from that proposed for \SrOne, in which there is a crossover from incoherent to coherent behavior at low temperature near the Fermi energy, i.e., a Fermi liquid regime. This is the type of behavior expected in a Hund metal picture, and here it is extended to a ferromagnetic system.


Other aspects of the data can also be mentioned.
The experimentally measured collapse of the spin polarization of the narrow bands in \SrFour\ occurs $\approx$15-20K higher than the bulk $T_c$.
A possible origin of such behavior is enhanced magnetism at the surface.
The most straightforward explanation for a surface-enhanced $T_c$ is the larger rotation angles of the surface RuO octahedra that generally occur in the layered ruthenates, resulting in predictions of a FM surface layer in \SrTwo\ \cite{Matzdorf2000} and recent STM experimental evidence for a controllable FM layer at the surface of nonmagnetic \SrThree\ \cite{Naritsuka2023}.
An enhanced octahedra rotation angle in the outer surface layer of a \SrFour\ has recently been predicted from a structually-relaxed surface slab DFT calculation \cite{Marques2024}.

Returning to the issue of localized signatures from decoherence broadening, in addition to the broad incoherent states discussed above that make much of band structure invisible between the O-bands and a few hundred meVs below \EF,  there is a clear asymmetry in the lifetime broadening versus binding energy  and as a function of temperature between spin-majority and spin-minority states.
Such a spin-dependent lifetime asymmetry is well understood to originate from the spin-splitting of bands near \EF\ causing a large asymmetry in relative fraction of occupied and unoccupied states for the two spin-channels, and thus an enhanced probability of spin-majority states spin-flip scattering into the greater number of unoccupied spin-minority final states than vice versa \cite{Monastra2002}.
This same underlying asymmetry of spin-flip scattering physics is also reflected in 
the larger spin-majority $T$-dependent energy shifts (of weakened Stoner exchange) of the Ru $d$ \EF-crossing bands in Figs. \ref{fig_tdepsp}(d)$-$\ref{fig_tdepsp}(f), as well as for the oxygen flat band in Fig. \ref{fig_tdepo}.

\subsection{Band renormalization and narrow band $T$-dependence} \label{DiscussNB_Tdep}

The dispersion of the bands near the Fermi energy measured at low temperatures and the dramatic T-dependences of the narrow band amplitudes show the large effects of interaction and correlation in this system. The magnitude of the effects are comparable to what has been found in
\SrTwo\ and \SrThree, which bring up enticing issues on the nature of the correlations. Those systems have long been regarded as the primary examples of Hund metals, which have been identified as a class of systems with multiple bands where exchange interactions lead to remarkable effects \cite{Georges2013,Medici2011,Deng2019}. This is in contrast (see especially Ref. \cite{Deng2019}) to Mott insulators epitomized by the single band Hubbard model where the large effects of correlation occur only if the on-site interaction is comparable to or larger than the band width \cite{Imada1998}. The most prominent classes of materials with such large local Coulomb interactions are the 3$d$ and 4$f$ transition metals, especially the oxides.  However, in the 4$d$ systems such as ruthenium compounds, no such large interactions are expected. The critical step in identifying the class of Hund metals was the realization that correlation could play such a large role in these materials because of the multiple bands near the Fermi energy and the atomic Hund's rule exchange interactions even if the magnitude of the interactions are smaller than the band widths. A characteristic feature is a Kondo-like effect, but instead of a single energy scale there are multiple scales for orbital and spin fluctuations.  The characteristic energies for orbital fluctuations are larger than for spins, leading to a two-stage process where there is a crossover from orbital and spin-fluctuation regimes to coherent bands only below a spin-fluctuation coherence temperature \cite{Kugler2020}.

Ferromagnetic \SrFour\ provides challenges and the opportunity to reveal new effects. As shown by the results of the present work, the bands are essentially two dimensional with dispersion similar to those in \SrTwo\ and \SrThree.   However, there is a large difference because of the splitting of the bands in the trilayers and splitting of the majority and minority spin bands which lead to very different occupations of the bands.  As mentioned above, the fact that the splitting of the majority and minority spin bands barely changes as a function of temperature, even above the transition temperature, is strong evidence of magnetism due to local moments that persist to high temperature even though they become disordered. 
A challenge  for the Hund metal picture is to understand what happens in the ferromagnetic state  where there are very different Fermi surfaces for majority and minority spin bands, and where there is spin order so that spin fluctuations are reduced below T$_c$ \cite{Dang2015}.

A hallmark of Hund metal correlations is an energy renormalization of bands close to \EF.
From the Table \ref{table1} summary of band velocities and effective masses with respect to the mean-field DFT calculations in Sec. \ref{DFT-comparison}, we observe both a moderate renormalization factor of 2 for the description of the narrow band energies and dispersions (after exchange splitting correction), and a larger 3$-$5$\times$ renormalized ``kink'' in the slope of the \EF-crossing bands. 
The larger renormalization factors are similar to those found experimentally for \SrTwo\ \cite{Tamai2019} and \SrThree\ \cite{Allan2013}, and in DFT+DMFT calculations of \SrTwo\ \cite{Tamai2019} and \SrOne\ \cite{Hahn2021} with Hund exchange show ``rounded'' band velocity renormalizations with similar overall factors. 
On the other hand, correspondence of the kink energy scales in \SrOne\ \cite{Yang2016} and in \SrFour\ \cite{Ngabonziza2020} to Raman spectroscopy vibrational modes have suggested an electron-boson coupling origin to ARPES dispersions described as having ``abrupt'' kinks.
The coexistence of both electron-phonon and electron correlation origins to the mass renormalization and their different low and high energy scales is also discussed for \SrTwo\ \cite{Iwasawa2020}  as a means for distinguishing the two effects.

A second hallmark of Hund metal correlations is the $T$-dependent coherence-incoherence crossover behavior that leads to a reduction in the quasiparticle coherent weight with increasing temperature, as well as temperature-dependent lifetime effects.
For quantifying the first aspect of coherent spectral weight, we integrate the area of the two narrow peaks in Figs. \ref{fig_tdep}(a) and \ref{fig_tdep}(b) after the background subtraction employed in Figs. \ref{fig_tdep}(c)$-$\ref{fig_tdep}(e) and Appendix Fig. \ref{fig_spindet}(c) for the correction to the spin-polarization. 
The assumption here is that we have isolated the quasiparticle peak from extrinsic energy loss and incoherent background contributions.
We note that the experimental resolution broadening effect is area-preserving and thus does not affect the relative $T$ dependence of the areas. 
The result, plotted in Fig. \ref{fig_thund} (a) is a normalized profile that exhibits a linear decline down to 40\% of the low $T$ area for the zone-centered ($\delta$) narrow band, and $\sim$50\% for the zone boundary ($\varepsilon$) narrow band.

We then compare this result to two experimental examples in the literature also with ruthenate Hund metal behavior inferred from $T$-dependent peaks: (i) a \SrTwo\ surface state \EF-crossing band  
\cite{Kondo2016} and (ii) a heavy bulk band in \CaOne\ \cite{Liu2018}. 
In both cases, the spectral weight $T$ dependencies also exhibit quasilinear profiles over a similar $T$ range as our \SrFour\ data sets \cite{Sr214fit}. 
While the \CaOne\ spectral weight reduction  is very similar to that of the two \SrFour\ narrow bands, the \SrTwo\ spectral weight completely disappears by 160 K.

\begin{figure}[ht]
\begin{center}
\includegraphics[width=8.5cm]{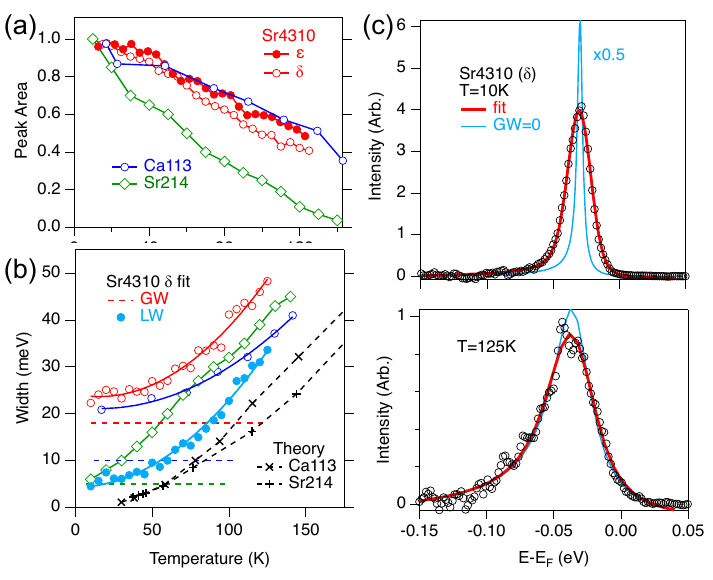}
\caption{
\textbf{Temperature dependent coherence crossover behavior comparison to theory.}
(a) Comparison of the background-subtracted peak area of the \SrFour\ ($\delta$) and ($\varepsilon$)  narrow peaks in comparison to two other studies of \CaOne \cite{Liu2018} and a \SrTwo\ surface state \cite{Kondo2016}.  All exhibit approximately linear reductions of the quasiparticle spectral weight. 
(b) The raw data $T$-dependent peak widths (open circles) of the three systems, as defined in (a), measured with different energy resolutions (dashed horizontal lines), in comparison to theoretical DMFT calculations of quasiparticle scattering rates (dashed line) for \SrTwo\ \cite{Mravlje2011} and CaRuO$_3$ \cite{Dang2015}.  Also compared are the intrinsic Lorentzian widths (LW) of the \SrFour\ peak (solid blue dots) extracted from lineshape fitting with a constant Gaussian width (GW) instrumental broadening.
(c) Example lineshape fits of the \SrFour\ low- and high-$T$ spectra to an asymmetric Lorentzian convolved with a Gaussian. Red lines are the overall fit and the blue lines are the intrinsic Lorentzian lineshape.
}
\label{fig_thund}
\end{center}
\end{figure}

For the second aspect of temperature-dependent lifetime effects, we next compare the raw full peak widths of the three ruthenates in Fig. \ref{fig_thund}(b). While the width of the \SrTwo\ surface state peak, measured with 5 meV resolution, varies rather linearly from from 6  to 40 meV \cite{Sr214fit}, the \CaOne\ and \SrFour\ peak widths exhibit nonlinear behavior from 20 to 40$-$50 meV, measured with 10 meV and 18 meV resolution (dashed horizontal lines), respectively.  
Next, we also make comparison to DFT+DMFT calculated $T$-dependent quasiparticle scattering rates ($\Gamma$) for \SrTwo\ \cite{Mravlje2011} and CaRuO$_3$ \cite{Dang2015}, extracted from plots of $\Gamma/T$.   Here we observe, overplotted in Fig. \ref{fig_thund}(b), predictions of few meV widths at low $T$  rising to 30 meV over the experimental $T$ range. 
The theory profiles are weakly quadratic up to $\sim$100K meV before flattening out to be more linear to higher $T$. 
Clearly the energy resolution limitations hamper the experiment-theory comparison, even for the 5 meV resolution measurement. 

To improve the comparison to theory for \SrFour, we perform quantitative lineshape fitting at each temperature to an asymmetric Voigt function with a fixed instrumental Gaussian convolution width. One advantage of the line-shape fitting of the large dynamic range \SrFour\ narrow peak ($\delta$) is its finite binding energy which allows fitting of the full peak profile with negligible influence of the $T$-dependent Fermi-Dirac distribution \EF-cutoff that enters in the \CaOne\ and \SrTwo\ surface state analyses.
Figure \ref{fig_thund}(c) shows example lineshape fits to the lowest 10K and highest 125K spectra. 
The low-$T$ fit requires a Gaussian-dominated lineshape to fit the weak tails, implying a much narrower intrinsic Lorentzian width, while the high-$T$ fit exhibits a more $\Lambda$-shaped Lorentzian profile implying a large increase of the intrinsic width to be much greater than the instrumental broadening. 
The resulting  intrinsic quasiparticle Lorentzian widths derived from the line-shape fitting are overplotted in Fig. \ref{fig_thund}(b), and indeed exhibit a large quadratic-like $T$ dependence from 5 to 35 meV.   
The general agreement with the Hund metal theory predictions is now much closer with an approximate constant offset that suggests of the need for the inclusion of a few meV constant-width impurity scattering contribution.

Additionally, Hund DMFT calculations predict strong orbital-dependence of the spin-screening coherence evolution due to the presence of enhanced scattering at a 2D van Hove singularity for one orbital \cite{Mravlje2011,Kugler2020}.
In the presence of magnetic order, spin-dependent lifetime asymmetries and the inherent suppression of spin-fluctuations, e.g., when the magnetization becomes fully saturated,   
are expected to give additional spin-dependent variation of $T$-evolutions of coherence and mass enhancement \cite{Dang2015}.
The system of \SrFour\ studied here combines all these aspects of Hund correlations plus ferromagnetism plus 2D van Hove singularities. While spin-dependent asymmetries have been discussed in the previous section for the \EF-crossing bands, it interesting to observe that a very similar $T$ dependence is observed for the spin-minority hole-like vHS band ($\delta$) at the zone center  and the spin-majority saddle-band vHS ($\varepsilon$) at the zone boundary.

\begin{figure*}[ht]
\begin{center}
\includegraphics[width=18cm]{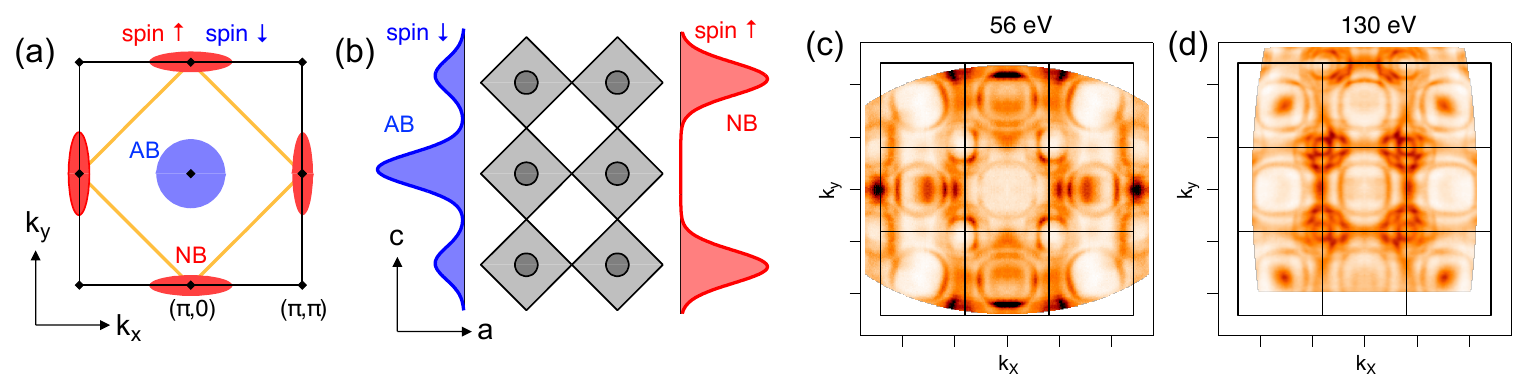}
\caption{
\textbf{Summary of low temperature \SrFour\ spin-polarized narrow bands.}
(a) Momentum-space schematic of the highly coherent -30 meV narrow spin-majority saddle-bands at (\p,0) and spin-minority hole-band at (0,0). Zone-folding effects are not represented.
(b) Schematic of the spatial distribution of the narrow band weights  in the outer and central layer of the trilayer structure.
(c,d) Multizone checkerboard-like enhancements of zone-center hole-band intensities consistent with a central layer-only body-centered site photoionization matrix element structure factor for two different photon energies.
}
\label{fig_summary}
\end{center}
\end{figure*}

\subsection{Layer-dependent magnetism} \label{DiscussLayer}

Layer-dependent spin and orbital moments in \SrFour\ have been modeled from polarized neutron diffraction analysis \cite{Granata2016,Forte2019}, and models of layer-dependent spin orientations have been proposed for the in-plane metamagnetic behavior \cite{Zhu2018,Capogna2020}.
The layer dependence has been generally discussed in terms of the structural distortion inequivalence of the Ru sites in the central and outer layers, e.g., different rotation angles or apical flattening (elongation) of the RuO$_6$ octahedra \cite{Forte2019}. Indeed, magneto-lattice coupling is evidenced by temperature-dependence lattice constants correlated to the magnetic behavior \cite{Granata2013} and by the application of modest pressure that induces a transition from $c$-axis ferromagnetism to a basal plane antiferromagnetic state \cite{Zheng2018}.

Here we identify an additional electronic spatial inequivalence of the narrow band states between the central and outer layers, whose origins are separate from the structural distortion considerations above.
As schematically shown Fig. \ref{fig_summary}(a), the two distinct high-DOS narrow band states below \EF\  exhibit large momentum-separation, opposite spin polarization,
and also possess different AB versus NB \dxzyz\ orbital origins from the comparison to DFT.
The NB orbital has an odd symmetry relative to the central layer mirror plane, and thus contains  a node in the central layer, while the B and AB orbitals have even symmetry and contain mixed layer character, but with a more dominant central layer spatial origin.
Hence, the high-DOS spin-minority narrow band at \G\ with AB band origin is predicted to be spatially localized primarily in the central layer, while the high DOS spin-majority saddle-point van Hove singularity at M with NB band origin is  localized exclusively  in the outer layers, as schematically represented in Fig. \ref{fig_summary}(b), and described in more detail in the Appendix \ref{appendix:matrix}.

Experimental evidence for this theoretical prediction of spatial localization of the narrow bands, exists in the multiple-BZ  $k$-space modulation of their ARPES intensities.
Consideration of only the central layer as a photoelectron emitter for the \G-point AB hole-band greatly simplifies the photoionization matrix element structure factor for that orbital.
The stacking of the trilayers forms the body-centered tetragonal lattice shown in  Fig. \ref{fig_intro}(a)
which, analogous to a \textit{bcc} lattice, results in maximal intensity
 for BZ indices ($hkl$) with ($h$+$k$+$l$) being even, and full suppression for ($h$+$k$+$l$) being odd.
 Neighboring BZs in momentum space, with a relative index change of one, will thus have opposite structure factor intensities.
Such a checkerboard pattern of high and low intensity at the zone-centers is experimentally observed and plotted in Fig. \ref{fig_summary}(c) and \ref{fig_summary}(d) for two-fold symmetrized maps at 56 and 130 eV, where the high/low intensities are reversed.
A matrix element orbital angular distribution term is responsible for the suppressed intensity in the first BZ for the 130 eV map (see Appendix \ref{appendix:matrix}).

The spatial localization of the NB \dxzyz\ states to the outer layers only is even more selective than for the AB states, and has implications for models of the in-plane metamagnetism as discussed in the next section.

\subsection{Zone-folding and band hybridization} \label{DiscussZonefolding}

Many of our insights into the \SrFour\ spin structure in comparison to DFT calculations are obtained
via simplification and neglect of the complexities of the oxygen octahedral rotations (and zone-folded band structure) and of the effects of SOC spin mixing.  This is justified from the experimental weakness of the zone-folded band structure effect at the photon energies used in the study.
Similarly, the DFT calculations including spin-orbit coupling predict less than 5\% spin-character mixing of the bands, and thus are consistent with the very high (but not 100\%) experimental spin polarization of the narrow bands.
Hence, these perturbative complexities do not affect any of our basic conclusions about narrow bands or layer-dependent magnetism.

One example of zone-folded band intensities too strong to ignore in our measurements is at the (3\p,0) M-point in Fig. \ref{fig_tdepsp} where a distinct hole-band ($\gamma^\prime$)  intercepts the strong intensity spin-majority saddle band ($\varepsilon$) and appears to not reach the Fermi level.
Clear evidence of a strong interaction between the spin-up and spin-down bands, enabled by the spin-mixing of the spin-orbit interaction, highlights a complexity to this region of the band structure.
Here we probed the $T$ dependence of this band crossing region
and revealed a progressive weakening of the hybridization due to the decoherence of the spin-majority band, which then reveals, in the thermally excited states above \EF, the two distinct bands coming from the spin-majority and spin-minority band dispersions above the hybridization gap. The two bands form a small electron dispersion with opposite-spin edges and a band minimum very close to \EF. At the highest measured temperature just above $T_c$, the hybridization is weak enough so that the rapidly dispersing spin-minority band does extrapolate directly to \EF, and its zone-folded FS contour becomes more visible.

  Recently high resolution ARPES using 21$-$32 eV photon excitation  has been able to visualize finer details of the low temperature hybridization gapping of this opposite-spin band crossing
 and correlate the near-\EF\ tips of the ($\varepsilon$) saddle band to meV-scale van Hove singularity dispersions observed by STM-QPI \cite{Marques2024}.
 Moreover, a tight-binding minimal model energy sensitivity of the vHS to $ab$-plane magnetic fields was proposed as a possible Lifshitz FS reconstruction origin of the field-dependent metamagnetic behavior of \SrFour.

The outer-layer spatial localization of the nonbonding zone-boundary narrow saddle-band involved in the creation of the meV-scale van Hove singularities, implies that such metamagnetic tuning would also occur solely in the outer layers.
Additional correlation of metamagnetism to the outer layers has also been demonstrated by a recent study of 4\% Ir-doped  \SrFour, in which an $\sim$1\deg\ enhanced \textit{outer} layer octahedral rotation angle is measured coincident with an increase in the in-plane metamagnetic critical field from 2.5 to 4.5 T \cite{Ye2023}.
The electron-doping of the system with Ir substitution for Ru should have a straightforward effect of shifting the energy of the vHS to higher binding energy, thus requiring a larger in-plane field to achieve the same Lifshitz transition tuning of the vHS through \EF.  Also contributing to the vHS energetics would be the enhanced spin-orbit coupling of the 5$d$ Ir states leading to larger hybridization gapping.  The general $T$-dependent decoherence of the spin-majority saddle-band and progressive weakening of the hybridization gapping to higher temperature, may contribute to the anomalous low-field susceptibility with maximum at $\sim$50K, and to the overall $T$-$H$ phase diagram \cite{Gupta2006}.

\section{Summary} \label{summary}

This work shows that  \SrFour\ displays a remarkable range of spin-polarized features
that extend our understanding of the series of strontium ruthenates.
The experimental findings have been presented in terms of an extensive study at low temperature and results for temperature dependence of certain features.
At low temperature the primary experimental results are:
(i) Distinct Fermi surfaces with opposite spin polarization are found at the zone center and the zone corner, with weak intensity oxygen-octahedron zone-folding effects.
(ii)  Two separate very high intensity narrow bands exist  30 meV below the Fermi-level, a hole-like band-maximum  vHS located at the zone-center and a saddle-point vHS 
at the zone boundary, also exhibit opposite nearly-pure spin polarization.
(iii) An intrinsic incoherent spin-majority background exists over the entire Ru-$d$ valence band region.
(iv) Sharp spin-minority oxygen bands exist at the top of the oxygen manifold.

Comparison of the sharp low $T$ FS contours and narrow band energies to spin-polarized DFT calculations, allows (i) quantitative evaluation of the predictions of different exchange-correlation functionals, and (ii) identifies the zone-center and zone-boundary narrow bands to have different antibonding versus nonbonding  \dxzyz\ orbital origins, which in turn have (iii) different spatially localized weight distributions in the central and outer layers.
(iv) The central-layer-dominant antibonding \G-point narrow band localization provides a natural body-centered structure factor explanation for the experimental multi-BZ selectivity of its ARPES intensities.
(v) The outer-layer-only localization of the nonbonding zone-boundary narrow saddle-band
indicates that the metamagnetic tuning model of meV-scale van Hove singularities \cite{Marques2024}, created by band-crossing hybridization of a zone-folded \dxy\ band with the zone-boundary narrow saddle-band, also occurs solely in the  outer layers.

The temperature-dependence of the electronic structure of \SrFour\ additionally reveals numerous features that are important for the understanding of the magnetic behavior.
(i) The narrow bands exhibit dramatic temperature-dependent amplitude suppression, with 60\% loss of integrated coherent spectral weight, upon warming to $T_c$. We attribute this to Hund metal coherence-incoherence crossover behavior \cite{Georges2013,Medici2011,Deng2019,Mravlje2011,Dang2015}, 
and provide a experiment-theory comparison of the $T$-dependent linewidth behavior.
(ii) The spin-resolved $T$ dependence of the narrow bands and the incoherent spin-majority background, both maintain a strong spin polarization up to $T_c$ similar to the bulk magnetization profile, and also exceed $T_c$, likely due to enhanced surface octahedral rotations.
(iii) The $T$ independence of the narrow band energies give clear evidence for Heisenberg localized behavior with a constant exchange energy up to $T_c$ and clear departure from the predictions of simple itinerant ferromagnetism.
(iv) Fermi-edge crossing bands centered at X and $\Gamma$, 
and the spin-polarized O-bands at M, exhibit spin-asymmetric $T$-dependent energy-shifts of $<$60 meV that are also far from a complete Stoner collapse, and spin-asymmetric enhanced decoherence broadening of the spin-majority bands.
(v) The $T$-dependent coherence of the zone-boundary saddle-band, affecting its hybridization and the energetics of the meV-scale van Hove singularities, may contribute to the anomalous magnetization $T$ profile associated with in-plane metamagnetism.

The strong coherence-like $T$-dependent amplitude, opposite spin polarization, momentum separation,  and layer-specific localization of the narrow band vHS states, as well as $T$-dependent hybridization with zone-folded bands, are all key new  electronic structure ingredients to be considered for the modeling of magnetic and metamagnetic behaviors in \SrFour.
The results and interpretations here may be relevant to the low-dimensional van der Waal ferromagnet \FGT, also a three-layer structure with two unique Fe sites, where coherence-like $T$-dependences  in ARPES \cite{Zhang2018} and optical conductivity \cite{Corasaniti2020} have been recently discussed in a theoretical framework of site- and orbital-dependent Hund metal correlations \cite{KimFGT2022}.

\section*{Acknowledgements}
We acknowledge fruitfall discussions with Changyoung Kim.
Research used resources of the Advanced Light Source, which is a US Department of Energy, Office of Science User Facility under contract no. DE-AC02-05CH11231.
This work was supported by the 2019 Innovation Fund of the American Physical Society through the U.S.-Africa Initiative in Electronic Structure. We gratefully acknowledge the African School for Electronic Structure Methods and Applications (ASESMA) and the Abdus Salam International Centre for Theoretical Physics without which this work would not have been done.
This work used the Centre for High-Performance Computing (CHPC), South Africa.
GC acknowledges NSF support via grant DMR-2204811.

\appendix
\setcounter{section}{0}
\renewcommand{\thesection}{\Alph{section}}       
\renewcommand{\thesubsection}{\Alph{subsection}}       

\section{Spin-resolved ARPES detection} \label{appendix:spindet}

Spin-resolved ARPES was measured at the Advanced Light Source beamline 10.0.1 using a Scienta R4000 spectrometer equipped with DA30 deflector plates for spin-integrated ARPES mapping and for steering electrons into dual very low energy electron diffraction (VLEED) spin-detectors.  
The two exchange-scattering type spin detectors use in-plane magnetization of FeO thin film targets to provide ($k_z,k_x$) and ($k_z,k_y$) components of the spin-asymmetry, e.g., with redundancy in the $k_z$ component.
For each spin-detector, the sequentially measured spectra $I_{+}(\omega)$ and $I_{-}(\omega)$ are used to compute the raw spin-scattering asymmetry, $A_{\pm}(\omega)= (I_{+}-I_{-})/(I_{+}+I_{-})$, which is corrected by the \textit{effective} instrumental spin-scattering efficiency factor, i.e., the Sherman function $S_\mathrm{eff}$, to determine the photoelectron spin-polarization $P(\omega)=A_{\pm}(\omega)/S_\mathrm{eff}$.
The corrected spin-dependent spectra are then calculated as $I_{\uparrow\downarrow}(\omega)=I_{av}(\omega)(1\pm P(\omega))$, where $I_{av}=(I_{+}+I_{-})/2$.
The effective spin-polarization correction factor can be viewed as a product of sample magnetization and detector scattering terms, $S_\mathrm{eff}$=$f \cdot S_\mathrm{det}$, where $f$ is the net fraction of the sample surface being probed with uniform domain alignment and $S_{det}$ is the exchange-scatting asymmetry factor of the detector.
The Sherman function for this exchange spin-detector is calibrated to be $S_\mathrm{det}$$\approx$0.25.

\begin{figure}[ht]
\begin{center}
\includegraphics[width=8.5cm]{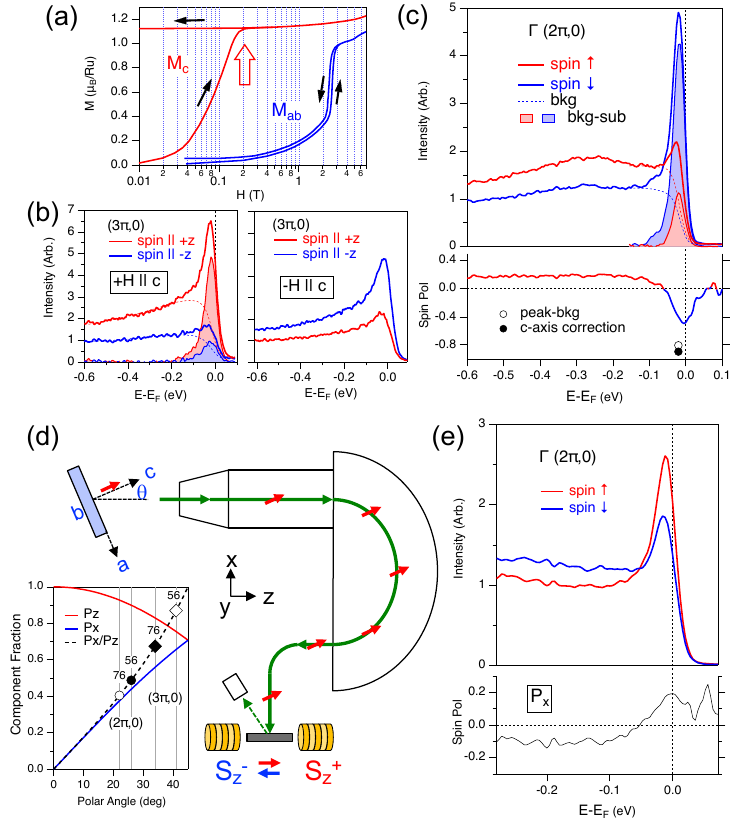}
\caption{
\textbf{Spin-detection procedure and analysis corrections.}
(a) Magnetization profile of \SrFour\ at $T=1.8$ K \cite{Cao2003} illustrating low temperature $c$-axis magnetization domain alignment with $\sim$0.2 T applied external field.
(b) Demonstration of the sign reversal of the experimental spin-spectra at (3\p,0) for a reversal of the $c$ axis external field.
[(b) and (c)] Example background subtractions of the two narrow band peaks,  resulting in enhanced spin-polarization asymmetry [open dot in (c)].
(d) Schematic of the $c$-axis spin-alignment relative to a $z$-axis spin-component detection, resulting in an additional 1/cos$\theta$ spin-polarization correction [solid dot in (c)].
(e) Nonzero $x$-axis spin polarization with magnitude (and sign reversal) consistent with the experimental polar angle difference between the spin-detection and a pure $c$ axis sample spin alignment.
}
\label{fig_spindet}
\end{center}
\end{figure}

To achieve uniform magnetic domain alignment for spin-resolved ARPES, a permanent magnet with $\approx$0.2 T field strength at the sample surface was used to field-cool the sample, both prior to and after sample cleaving.
The magnetization profiles of \SrFour\ in Fig. \ref{fig_spindet}(a) shows how such an external field strength parallel to the $c$ axis enables magnetization saturation (domain alignment) of the sample (even without field cooling) and how the magnetization is maintained when the external field is removed.
In contrast, the high-moment state for $ab$-plane magnetization requires a higher critical field and is not maintained at low temperature.
For this study, only a field alignment parallel (or antiparallel) to the crystalline $c$ axis was used, and only the $k_z$ components of the spin-asymmetry parallel to this magnetization axis are presented.
Figure \ref{fig_spindet}(b) demonstrates how the sign of the detected spin-asymmetry is reversed if the sample is remagnetized (also with field cooling) with the external field reversed to be antiparallel to the $c$ axis.
The peak amplitude reduction reflects some surface adsorption aging during the thermal cycling.

The spin polarization of coherent peaks in the spin-dependent spectra requires additional modeling and subtraction of spin-dependent incoherent background profiles.   
Figures \ref{fig_spindet}(b) and \ref{fig_spindet}(c) show examples of such a procedure to separate out first the coherent peaks (shaded) of the low $T$ narrow band spectra, whose amplitudes  are then used to compute a single spin-polarization value.
Shirley-type integral background profiles are employed and an additional exponential decay is used to account for long incoherent character tails of the (3\p,0) narrow band spectra, that are perhaps related to its saddle-band orthogonal hole-like dispersion to higher binding energy.  An enhancement of the experimental spin polarization from 50\%\ to 80\% is observed for the (2\p,0) example (open dot).
The spin polarization of the (3\p,0) narrow peak is similarly enhanced to $>$80\%, with subtraction of its same-sign majority-spin background.
Similar background subtraction is employed for the spin-integrated narrow band spectra  for the evaluation of the quasiparticle peak area and $T$-dependent spectral weight loss profiles in Fig. \ref{fig_thund}(a).

A second quantitative spin-polarization correction arises from the large $\sim$26$^{\circ}$-34$^{\circ}$ sample polar angles (rotation about the sample vertical $b$-axis) required to measure at the second BZ (2\p,0) and (3\p,0) points at 56 and 76 eV, respectively.
The $P_z$ spin-polarization detection axis is not purely aligned to the sample magnetization axis, as shown in Fig. \ref{fig_spindet}(d), and thus contains a non-zero in-plane component, i.e. $P_z=P_c cos\theta + P_a sin\theta$.  Conversely, a $P_x$ spin polarization measurement could contain a non-zero $c$ axis component, i.e. $P_x=-P_c sin\theta + P_a cos\theta$.
For a pure $c$ axis sample spin-alignment ($P_a$=$P_b$=0),  $P_z$ should be corrected by $P_c=P_z/cos\theta$  to obtain the true $c$ axis spin polarization, e.g. $P_c$=1.11$P_z\sim$90\% for the (2\p,0) coherent peak in Fig. \ref{fig_spindet}(c).
Also a nonzero spin polarization of $P_x = -P_c sin\theta = -P_z tan\theta$ is expected, where the negative sign is dependent on the geometry and definition of the in-plane magnetizing coils.
 Indeed, Fig. \ref{fig_spindet}(e) experimentally shows $P_x$ spin-resolved spectra for the (2\p,0) point narrow band exhibiting $\sim$50\% smaller spin-polarization than the raw $P_z$ component consistent with $tan(26^{\circ}$) = 0.49.
The above analysis of $P_z$ and $P_x$ spin-polarization components indicating $P_c$ greater than 90\% at the intense narrow band $k$-points, puts some constraints on, but does not preclude models of a small amount of canting of the net spin relative to the $c$ axis \cite{Gupta2006}.

\section{Spin-integrated ARPES photon dependence} \label{appendix:photon}

Spin-integrated photon-dependent mapping of the electronic structure, both at normal emission and off-normal, was key to the identification of the $k$-space locations at which the narrow flat bands at \G\ and at the tetragonal Brillouin zone (BZ) boundary saddle-point were strongest intensity and most suitable for subsequent spin-resolved measurements.
Figure \ref{fig_hv}(a) shows a normal emission (first BZ) \EF-intensity map for LH polarization in which $no$ photon energy in the range of 30$-$124 eV exhibits a strong matrix element enhancement of  the \G-point flat band.
In contrast for LV polarization photon-dependent mapping, a strong enhancement centered on 76 eV is observed in Fig. \ref{fig_hv}(b).  However, since LV polarization was not available at the spin-ARPES beamline 10.0.1, further angle-dependent mapping into the second BZ was explored.

\begin{figure}[ht]
\begin{center}
\includegraphics[width=8.5cm]{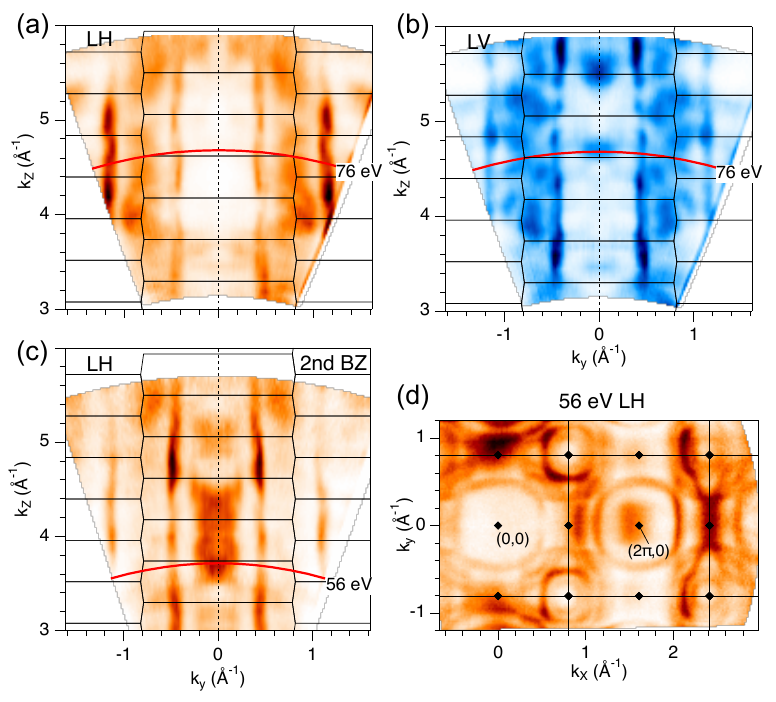}
\caption{
\textbf{Photon energy dependent $k_x$-$k_z$ maps.}
(a) Normal emission map using LH polarization.
(b) Normal emission map using LV polarization.
(c) Off-normal second BZ map using LH polarization.
(d) Angle-dependent FS map at 56 eV using LH polarization.
}
\label{fig_hv}
\end{center}
\end{figure}

Indeed at 76 eV and using LH polarization for angle-dependent mapping, a strong enhancement of the (2\p,0) \G-point narrow band intensity is observed in Fig. \ref{fig_intro} of the main text.  In addition, a strong intensity enhancement of the (3\p,0) saddle-point region, as well as the large square FS contour at ($\sim$2.5\p,0), are observed and used for the spin-ARPES measurements.
To further explore the relative (2\p,0) matrix element enhancement, an off-normal photon-dependent map with a constant $k_{\vert\vert}$= 1.62 \AA$^{-1}$ centered on the second BZ, was performed. As shown in Fig. \ref{fig_hv}(c), an even greater enhancement was discovered at 56 eV, using LH polarization, which also coincidently corresponded to the maximum flux performance of the spin-ARPES beamline.
Thus 56 eV could be chosen to optimize the (2\p,0) \G-point spin-ARPES measurements without having to compromise the energy resolution settings that are very comparable to those at ALS beamline 4.0.3.
Figure \ref{fig_hv}(d) shows the corresponding angular dependent $k_x$-$k_z$ FS map at 56 eV showing the (2\p,0) second BZ center enhancement of the \G\ flat band.

\section{Exchange functional comparison} \label{appendix:dft}

In Fig. \ref{fig_dftfunc}, we provide a side-by-side comparison of three spin-polarized DFT calculations with different nominal exchange splittings of approximately 0.6, 0.9 and 1.1 eV for the key relevant bands close to the Fermi level.  (Note that there is a variation of the exchange splittings for different bands as illustrated in Fig. \ref{fig_dftfunc}.)
The PBEsol and PBEsol+$U$ (1 eV) bands are taken from Ref. \cite{Gebreyesus2022} and replotted with left/right separation of spin-up/down states for greater clarity. The octahedral rotation and zone-folding effect, indicated by the (0,0) and (\p,\p) labeling equivalence,  has not been unfolded to the tetragonal BZ [as performed in Fig. \ref{fig_spindft}(a)].
The LDA functional calculation, new to this work, is similarly plotted.
A recent PBE96 calculation from Benedi{\v{c}}i{\v{c}} \etal\ \cite{Benedicic2022} also predicts a large exchange splitting value of $\sim$1.1 eV similar to the PBEsol+$U$ calculation, in which the Fermi-level cuts just slightly below the highest energy (\p,\p) hole band.

The better agreement of the LDA functional to experiment, the systematic trend towards increasing exchange splitting and larger moments for LDA $\rightarrow$ PBEsol $\rightarrow$ PBE96 $\rightarrow$ DFT+$U$, and the prediction of a large moment half-metal  groundstate for small $U$ is consistent with previous comparative studies of DFT exchange-correlation functionals for magnetism in \SrOne\ \cite{Granas2014}, \SrThree\ \cite{Rivero2017} and FM transition metals \cite{FuSingh2019}.

\begin{figure*}[ht]
\begin{center}
\includegraphics[width=16.5cm]{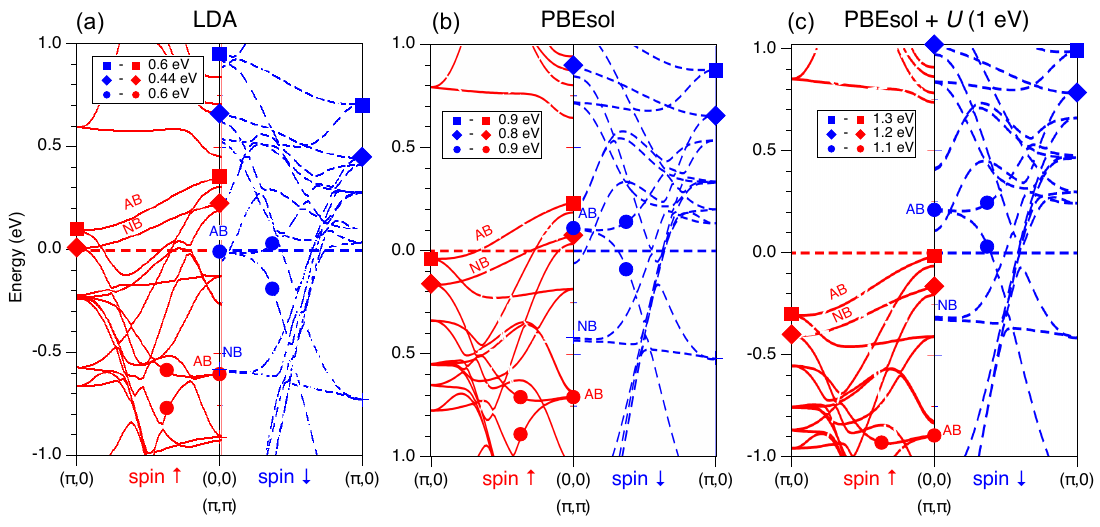}
\caption{
\textbf{Comparison of different exchange-correlation functionals}
Spin-polarized \SrFour\ for three different functionals of (a) LDA (b) PBEsol and (c) PBEsol + $U$ (1 eV).  The \EF\ result in (c) is very similar to a recent PBE96 calculation \cite{Benedicic2022}. Round, diamond and square symbols help to identify key equivalent spin-up and spin-down bands for purpose of evaluating the  band-dependent exchange splitting variations quantified in the inset boxes.
}
\label{fig_dftfunc}
\end{center}
\end{figure*}

The comparison in Fig. \ref{fig_dftfunc} also highlights the main text logic flow for the variable exchange-splitting exercise to find a consistent assignment of both the $-$30 meV narrow hole band at (0,0) and the saddle-point band at (\p,0), as well as find a quantitative agreement with the (\p,\p) hole FS contours.

(i) The large DFT+$U$ exchange splitting result in Fig. \ref{fig_dftfunc}(c) of a half-metal in which the spin-majority states are completely filled is trivially ruled out due to the distinct experimental observation of spin-majority FS contours in the (\p,\p) region.  This also rules out an NB \dxzyz\ flat band origin for the narrow hole-like state at (0,0) which would require an even larger upwards spin-down energy shift and/or extremely large $>$10$\times$ energy renormalizations to achieve a $-$30 meV energy.

(ii) With the spin-minority AB flat state at (0,0) being the target assignment of the strong ARPES at (0,0), the PBEsol calculation  to PBEsol calculation in Fig. \ref{fig_dftfunc}(c) is observed to incorrectly predict the spin-minority AB state to reside just $above$ \EF.
 Also, while a shallow AB-derived spin-majority saddle-point band  is correctly predicted to be below \EF\ at (\p,0), it also predicts a second parallel dispersing NB-derived saddle-point band  to be just below it.  In contradiction, ARPES shows only \textit{one} $-$30 meV saddle-band dispersion.

(iii) A reduced overall exchange-splitting is proposed that both lowers the spin-minority AB state to be below \EF, and raises the spin-majority states to produce only a single NB-derived saddle-point dispersion below \EF.
Rigid shift corrections of +0.10 ($-$0.16) eV to the PBEsol spin-up (down) bands are selected, i.e. total exchange splitting of 0.64 eV ($\approx$0.9-0.26).    The spin-up correction is fine-tuned to provide a quantitative match to the three (flower + two nested square) FS contours at (\p,\p).
The large velocity spin-minority \dxy\ electron bands that form the large squarish contour(s) centered on (0,0) are less sensitive to the small corrections to the exchange splitting. Also consistent with ARPES, another \dxzyz\ NB-derived light mass band converges with the \dxy\ band near \EF.

(iv) The LDA functional calculation in Fig. \ref{fig_dftfunc}(a) also predicts a smaller Fermi-energy exchange splitting of  $\sim$0.6 eV in closer agreement to the ARPES-matching PBEsol exchange-splitting adjustment.
A rigid shift correction of $-$0.05 ($-$0.05) eV to the LDA spin-up (down) bands is required to obtain a similar result to the PBEsol exchange-splitting correction.

(v) The PBEsol and LDA calculations without spin-orbit coupling (SOC) predict two-fold band degeneracies of the B, NB and AB \dxzyz\ states at (0,0).  When the degenerate AB state is shifted to be below \EF\ to obtain agreement with the experimental shallow hole band at (0,0), this results in the prediction of an upwards dispersing electron-like \EF-crossing that is not observed experimentally.
This discrepancy is cured by the inclusion of SOC which splits this degeneracy at (0,0) as illustrated in Appendix \ref{appendix:SOC}.

(vi) The main text  discusses the comparison of the average magnetic moment ($M_{tot}$/3) of the PBEsol and LDA calculations to the experimental \SrFour\ low temperature saturated moment of 1.1 $\mu_B$/Ru for low-field $c$ axis magnetization \cite{Cao2003}. The moment increases to 1.2 $\mu_B$/Ru at higher applied fields of 7 T.
Larger experimental total moments up to 1.36 $\mu_B$/Ru have been reported \cite{Xu2007}, but are likely the result of contamination by ingrowth of \SrOne-like higher-order ruthenate layers \cite{Carleschi2014} which giving an additional $M$$\sim$0.2 contribution extending above T$_c$=105 K up to 160 K, i.e. the Curie temperature for \SrOne.
Table \ref{tablefct} summarizes these moment and exchange splitting values.

\setlength\abovecaptionskip{+0.3\baselineskip}
\setlength\belowcaptionskip{+0.3\baselineskip}
\sectionmark{Table Appendix. Exchange splitting and Moment summary }
{\setlength{\tabcolsep}{0.1em}
\begin{table}[h]
\caption{Comparison of experimental and spin-polarized DFT calculated exchange splittings and average magnetic moments.}
\begin{tabular}{c cc c cccc l}
 \toprule
 Exchange-Correlation & Exchange & Average & &  \\
Functional & Splitting & Moment$^a$
 & Reference  \\
 & (eV) & ($\mu_B$/Ru) & &  \\
 \colrule
     LDA (unrelaxed) & 0.44-0.6 & 1.36 & -    \\
     LDA (relaxed) & 0.44-0.6 & 1.17 & -   \\
     PBEsol  & 0.8-0.9  &  1.77 & \cite{Gebreyesus2022}  \\
     PBE96 & 1.1 & -$^b$ & \cite{Benedicic2022} \\   
     PBEsol+$U$ (1 eV) & 1.1-1.3 & $\sim$2 & \cite{Gebreyesus2022}  \\
     LDA+$U$ (2 eV) & 1.1-1.3 & $\sim$2 & -   \\
     \hline
     Energy-shifted PBEsol  & 0.64 & - & $\approx$ARPES \\ 
     Experiment, B$||$c, 0.1T  &- &  1.1 & \cite{Cao2003}  \\
     Experiment, B$||$c, 7T  &- &  1.2 & \cite{Cao2003}  \\
\botrule
\footnotesize{$^a$$M_{av}$=$M_{tot}$/3}.   
\footnotesize{$^b$\textit{value not reported.}}
\end{tabular}
\label{tablefct}
\end{table}
}

\section{Spin-orbit coupling}    \label{appendix:SOC}

We have ignored spin-orbit coupling (SOC) in much of the analysis since it is mainly a small effects and it complicates the analysis in terms of majority- and minority-spins.  However, it is important for details of the bands near the Fermi energy.
The DFT PBEsol calculations without SOC presented in Fig. 3(b) predict two extraneous  features (highlighed by yellow arrows) that are not experimentally observed.
(i) The PBEsol predicts two-fold band degeneracies of the B, NB, and AB \dxzyz\ states at (0,0), which result in the prediction of both downwards and upwards dispersing branches from the AB state.  Thus, after correcting the AB to be below \EF, an electron-like \EF-crossing is predicted which is not observed experimentally.
(ii) In addition, the bottom of spin-minority states at (\p,0) are only $\sim$20 meV higher than the (0,0) AB state energy. Thus the AB energy correction inevitably also shifts these spin-minority states to be crossing below \EF\ at (\p,0) -- which are also not experimentally observed.

\setlength\abovecaptionskip{-0.3\baselineskip}
\setlength\belowcaptionskip{-0.7\baselineskip}
\begin{figure}[ht]
\begin{center}
\includegraphics[width=8.5cm]{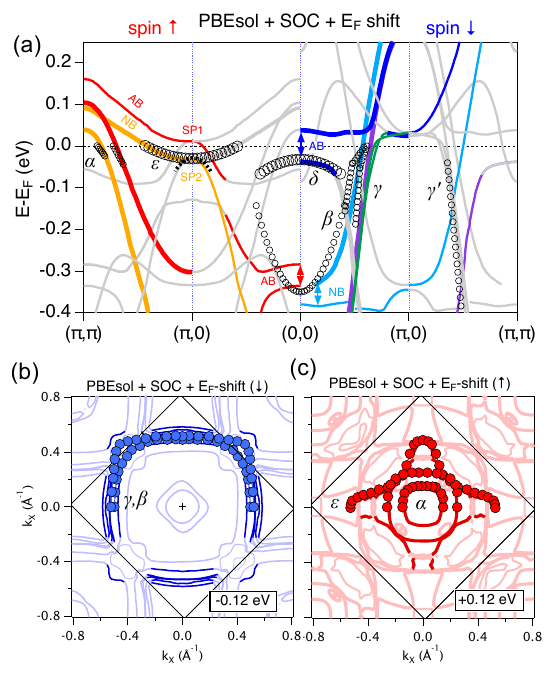}
\caption{
\textbf{PBEsol + spin-orbit coupling comparison to ARPES.}
(a) Energy-shifted and scaled PBEsol+SOC bands with $\approx$0.6 eV exchange splitting, giving similar (\p,\p) and (\p,0) spin-majority agreement to ARPES as in Fig. 3(b), but with key SOC-splitting of the spin-minority AB band degeneracy at (0,0).
(b) Experimental spin-minority FS contours ($\gamma/\beta$) compared to energy-shifted PBEsol+SOC FS contours. Lighter shaded contours correspond to zone-folded bands.
(c) Experimental spin-majority FS contours ($\alpha,\varepsilon$) compared to energy-shifted PBEsol+SOC FS contours.   Many of the lighter shaded contours correspond to spin-minority bands (with incorrect energy shifts) due to finite spin-mixing in the SOC calculation.
}
\label{fig_soc}
\end{center}
\end{figure}

 As shown by the DFT calculations \cite{Gebreyesus2022}, SOC splits the \dxzyz\ state degeneracies at (0,0) and creates two separated electron-like and hole-like narrow AB bands with even flatter band dispersions near (0,0).
Reduction  of the exchange splitting so that \EF\ lies in this spin-minority SOC gap, as shown in Fig. \ref{fig_soc}(a), preserves the AB energy level and renormalized dispersion agreement to ARPES, while eliminating the unobserved electron branch.
In addition, SOC pushes the bottom of the spin-majority states at (\p,0) to higher energy above \EF\ in alignment with the upper AB branch at (0,0).  This also cures the PBEsol presence of spin-minority FS contours in the (\p,0) region of Fig. 3(c), which are now absent in Fig. \ref{fig_soc}(b).
A symmetric rigid shift correction +0.12 ($-$0.12) eV to the PBEsol+SOC spin-up (down) bands also gives similar agreement as PBEsol (without SOC) to the (\p,\p) region FS contours [Fig. \ref{fig_soc}(c)].

A final observation from both the shifted and scaled PBEsol  plot in Fig.  3(b) and the PBEsol+SOC plot in Fig. \ref{fig_soc}(a) is that the lack of experimental spin polarization of the bottom of the experimental $\beta$ band at (0,0), i.e. the 0.2 eV wide spectral hump centered at $-$0.35 eV in Fig. 2(a), originates from the coincidental energy alignment of downwards-shifted spin-majority AB states and upwards-shifted spin-minority NB states.  The SOC splitting of these states contributes to the broadened width of the experimental spectral hump.

\section{Matrix element structure factor effects} \label{appendix:matrix}

The strong intensity variations of the \G-point flat band presented in the previous section originate from photoemission matrix element effects, which for a single orbital can be separated into a structure factor term multiplied by an atomic orbital angular distribution term \cite{Daimon1995}.
The angular distribution term is responsible for the strong photon-polarization dependence (LH versus LV) of the \G-point orbital, which is determined from the comparison to spin-polarized DFT
to be an antibonding \dxzyz\ orbital. The angular distribution term is also responsible for the largely suppressed intensity in the first BZ for LH-polarization for all photon energies.
The photoemission structure factor term originates from the interference effect of atomic orbitals with different spatial positions in the unit cell, and is responsible for the strong photon-dependent intensity modulations in the first BZ for LV-polarization, as well as the strong second-BZ selectivity of \G-point intensity for LH-polarization.

Since the above matrix element enhancements do not originate from any resonant photoemission process, e.g., where the photon energy is tuned to a core-level absorption threshold, we regard the enhanced low temperature spectral weight of the narrow bands at (0,0) and (\p,0)-equivalent points to be reflective of underlying high densities of states.
Instead, it is matrix-element orbital selection rules and/or structure factor phase cancellation that causes the \textit{suppression} of the spectral weight of the narrows bands at other (0,0) and (\p,0)-equivalent points.

\begin{figure}[ht]
\begin{center}
\includegraphics[width=8.0cm]{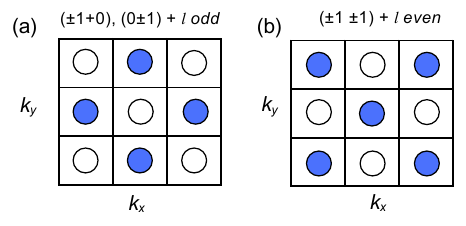}
\caption{
\textbf{Body-centered matrix element structure factor.}
Schematic checkerboard predictions of the body-centered
structure factor for the first and second Brillouin zone intensities being maximal (filled circle) for two different cases of BZ indices $\{h,k,l\}$ of
(a) $(h+k)$ odd with $l$ odd, and (b) $(h+k)$ even with $l$ even.
}
\label{fig_matrix}
\end{center}
\end{figure}

The triple-layer structure of \SrFour, with a half-unit cell lateral shift between triple-layers, results in six different Ru-sites per the unit cell and a complex photoemission structure factor for Ru-d photoelectron emission including all sites.
For a three layer system with a central layer and two equivalent outer layers (O-C-O), the eigenfunction solution can be understood by writing the 3$\times$3 matrix in terms of the basis of two even and one odd states: C, (O+O), (O-O). The odd state is decoupled and hence nonbonding (NB), and the even states define a 2$\times$2 matrix, with bonding (B) and antibonding (AB) solutions.
The odd symmetry NB eigenfunction solution is required to have no contribution from the central layer, and is spatially distributed with 50\% weight in each of the outer layers.
The B and AB eigenfunctions have weight on all the layers. If the central and outer layers are equal, there is 50\% in the central layer and 25\% in each of the outer layers;
in general, the ratios depend on the relative energies of the layers.
For a nonmagnetic nonrotated \SrFour\ structure, the DFT-calculated site-specific Ru \dxzyz\ orbital projections confirm the exclusive outer layer localization of the NB bands, and predict an even greater 3:1 ratio of central to outer layer contribution to the AB orbitals [as schematically illustrated in Fig. \ref{fig_summary}(b)].

As described in the main text and shown in Fig. 3, the analysis of the spin-polarized exchange splittings, identifies the \G-point flat band to have a specific AB \dxzyz\ orbital origin, and thus a dominant localization in the central layer.
Consideration of emission only from the central layers greatly simplifies the photoemission structure factor for the \G-point flat band to that of two lattice sites with body-centered
coordinates of $\vec{r}_1$=(0,0,0) and $\vec{r}_2$=($\frac{1}{2}$,$\frac{1}{2}$,$\frac{1}{2}$).
This is identical to the simplest textbook example of the body-centered cubic ($bcc$) structure factor of $F_{hkl} = f(1+e^{-i\pi(h+k+l)}) = f( 1+(-1)^{h+k+l} )$, where \{$h,k,l$\} are reciprocal space BZ indices.
The structure factor is maximal for $(h+k+l)$ being even and fully suppressed for $(h+k+l)$ being odd.
Thus, neighboring BZs in momentum space, with a relative index change of one, will have opposite structure factor intensity resulting in a checkerboard pattern of high and low intensity [as schematically illustrated in Figs. \ref{fig_matrix}(a) and \ref{fig_matrix}(b)].

The wide angle maps of \SrFour\ shown in Figs. \ref{fig_summary}(c) and \ref{fig_summary}(d) spanning out to \{$h,k,l$\}=$(\pm1,\pm1,l)$ (with data symmetrization) for two photon energies of 56 and 130 eV for LH-polarization.
The BZ selectivity of enhanced intensity for \{$h,k,l$\}=$(\pm1,0,l)$ or $(0, \pm1,l)$ where $l\approx9$ for 56 eV, is what is being exploited for this spin-ARPES study of the \G-point flat band at (2\p,0).
For 130 eV, the checker board pattern is reversed to be maximal for \{$h,k,l'$\}=$(\pm1,\pm1,l')$ or $(\pm1, \mp1,l')$ where $l'\approx10.5$. The first BZ intensity, however, is still suppressed from the atomic orbital matrix element term.
This photon energy dependent reversal of the checkerboard intensity patterns provides additional evidence of the operative body-centered structure factor, which in turn provides experimental proof of the central-layer spatial localization of the \G-point AB Ru-\dxzyz\ spin-minority flat hole-band.

\bibliographystyle{apsrev4-2_title}
\bibliography{../sro}


\end{document}